\documentclass[a4,12pt]{article}

\usepackage[utf8x]{inputenc}
\usepackage{ucs}
\usepackage{xcolor} %%%%% out
\usepackage[numbers]{natbib}
\usepackage{amsmath}
\usepackage{amsfonts}
\usepackage{amssymb}
\usepackage{multirow}
\usepackage{latexsym}
\usepackage{graphicx}
\usepackage{lineno}

\usepackage[position=bottom]{subfig}
\usepackage[margin=2.0cm]{geometry}
\usepackage{enumitem}
\usepackage[title]{appendix}
\usepackage{hyperref}
\hypersetup{
    colorlinks = true,
    urlcolor   = black,
    citecolor  = black,
}
\usepackage{xcolor}
\usepackage{tikz,siunitx}
\usetikzlibrary{shapes.geometric,shapes.symbols}
\usepackage{etoolbox}
\usepackage{multirow}
\usepackage{subfig}

\definecolor{myblue}{rgb}{0 0 1}
\definecolor{myred}{rgb}{0.7412 0.2588 0.2588}
\definecolor{mypurple}{rgb}{0.2929 0 0.6914}
\definecolor{mypink}{rgb}{1.0 0 0.5859}

\usepackage{soul, xcolor}
\newrobustcmd*{\hexagram}[1]{\tikz{\draw[thick, draw=#1, fill=white] (0cm, 0cm) -- (0.02887cm, -0.05cm) -- (0.08661cm, -0.05cm) -- (0.05723cm, -0.1cm) -- (0.08661cm, -0.15cm) -- (0.02887cm, -0.15cm) -- (0cm, -0.2cm) -- (-0.02887cm, -0.15cm) -- (-0.08661, -0.15) -- (-0.05723, -0.1) -- (-0.08661, -0.05) -- (-0.02887, -0.05) -- cycle;}}
\newrobustcmd*{\emptytriangle}[1]{\tikz{\fill[mypink, draw=#1] (0,0)  -- (0.2,0.0cm) -- (0.1, 0.2) -- cycle;}}
\newrobustcmd*{\emptydiamond}[1]{\tikz{\draw[thick, draw=#1] (0,0) --
(0.0684cm,-0.103cm) -- (0cm, -0.205cm) -- (-0.0684cm,-0.103cm) -- cycle;}}
\newrobustcmd*{\emptysquare}[1]{\tikz{\fill[mypurple, draw=#1] (0,0) --
(0.2,0) -- (0.2cm, 0.2cm) -- (0,0.2cm) -- cycle;}}

    \makeatletter
\def\@fnsymbol#1{\ensuremath{\ifcase#1\or \dagger\or \ddagger\or
   \mathsection\or \mathparagraph\or \|\or **\or \dagger\dagger
   \or \ddagger\ddagger \else\@ctrerr\fi}}
    \makeatother
\begin{document}

\normalsize	
\author
{Paria Makaremi-Esfarjani%
  \thanks{Email address for correspondence:
    paria.makaremiesfarjani@mail.mcgill.ca}$\,$, Andrew J. Higgins\thanks{Email address for correspondence:
    andrew.higgins@mcgill.ca} \\
  Department of Mechanical Engineering, McGill University,\\ Montr\'{e}al, Qu\'{e}bec, Canada}

\date{}

\title{Magneto-Rayleigh--Taylor instability and feedthrough in a resistive liquid-metal liner of a finite thickness}

{\let\newpage\relax\maketitle}

\begin{abstract}
%%%
\small
The effect of magnetic tension and diffusion on the perturbation growth of a liquid-metal liner subjected to the magneto-Rayleigh--Taylor (MRT) instability is investigated. An initially magnetic-field-free liquid-metal slab of finite thickness is surrounded by two lower-density regions. Within the lower region, a constant axial magnetic field of arbitrary magnitude is applied. The numerical examination of the MRT instability growth, initiated by a seeded perturbation parallel to the magnetic field at the liner's unstable interface, is performed for both perfectly conductive and resistive liners. To this end, a novel level set-based two-phase incompressible solver for ideal/resistive magnetohydrodynamic (MHD) flows within the finite-difference framework is introduced. Utilizing the implemented numerical toolkit, the impact of different Alfv\'en numbers and magnetic Reynolds numbers on the MRT growth rate and feedthrough at the upper interface of the liner is studied. Accounting for the finite resistivity of the liner results in an increase in the MRT growth and feedthrough compared to the ideal MHD case. The results indicate that magnetic diffusion primarily affects the MRT growth rate for higher wavenumbers, while for smaller wavenumbers, the effect of finite resistivity is only observed over a longer duration of instability development. We further demonstrate that decreasing the Alfv\'en number results in the faster emergence of the magnetic diffusion effect on the MRT growth rate. It is also observed that a greater electrical conductivity jump across the liner results in an increased perturbation growth. Lastly, the impact of surface tension on MRT instability growth for both ideal and resistive MHD cases is studied across different wavenumbers, specifically for Bond numbers related to fusion applications.
%%%
\end{abstract}

\maketitle

\section{\label{sec:introduction} Introduction }
The Rayleigh--Taylor (RT) instability emerges when a lighter fluid undergoes an acceleration towards a denser fluid~\citep{Rayleigh1882, Taylor1950}. In a broader context, RT instability arises when opposing density and pressure gradients coexist, a condition that can be expressed mathematically as $\nabla \rho \cdot \nabla p<0$. When perturbations are of small amplitude, the RT instability leads to an exponential growth of these perturbations over time at an unstable interface. The rate of perturbation growth $(\omega)$ is contingent upon factors such as interface acceleration $(a)$, the relative densities of the two fluids $(\rho_\mathrm{light},\, \rho_\mathrm{heavy})$, and the wavenumber of the initial perturbation $(k)$, as depicted by the following equation~\citep{Taylor1950}
\begin{equation}
    \omega = \left(\frac{\rho_\mathrm{heavy} - \rho_\mathrm{light}}{\rho_\mathrm{heavy} + \rho_\mathrm{light}}\, k\, a \right)^{\frac{1}{2}}.
\end{equation}
Therefore, if the perturbation amplitude at time $t=0$ is denoted as $\xi_0$, the perturbation growth over time within the linear regime is expressed as $\xi(t) \propto \xi_0 e^{\omega t}$.

The introduction of a magnetic field can profoundly influence the growth of RT instability. Consequently, the magneto-Rayleigh--Taylor (MRT) instability has received great attention in the literature. While the mathematical representation of the RT instability, $\nabla \rho \cdot \nabla p<0$, remains applicable to the MRT case, the pressure term should be modified to also include the effect of the magnetic pressure~\citep{weis2015thesis}. The growth of the MRT instability is a significant concern for fusion concepts, specifically in scenarios involving a liner implosion~\citep{weis2014temporal}, such as inertial confinement fusion (ICF), magnetized liner inertial fusion (MagLIF), and magnetized target fusion (MTF). In the context of ICF, the process involves compressing a cryogenic deuterium-tritium target by employing laser-driven ablation of a thin shell within a nanosecond time frame. The acceleration stage, driven by laser-ablated material, induces MRT-driven perturbation growth at the shell's outer surface. These perturbations can propagate to the inner surface and experience further amplification around the point of maximum compression, where the inner surface of the shell is decelerated by the lighter deuterium-tritium target~\citep{wang2015weakly,huneault2019rotational}. MagLIF is a magnetically driven ICF approach presently being developed by Sandia National Laboratories, in which a pre-heated and pre-magnetized plasma is compressed to fusion conditions through the compression of an imploding shell of finite thickness~\citep{sefkow2014design, gomez2015demonstration}. This approach is based on the idea proposed by Linhart~\cite{linhart1961dynamic} and Harris~\cite{harris1962} to utilize an imploding conducting shell to increase the magnetic field density and reach thermonuclear fusion conditions.

Initially proposed in the 1970s Linus project, MTF is an alternative fusion approach which focuses on achieving fusion conditions by compressing plasma using a conductive imploding surface~\citep{turchi1980, brunelli2013}. This concept employs a mechanically collapsed cylindrical or spherical rotating liquid shell to compress the plasma target. The shell absorbs the ensuing energy released from the fusion reaction in the form of heat and kinetic energy, which is subsequently extracted from the liquid through a heat exchanger. A critical factor for successful fusion in this approach is maintaining stability at the plasma/liquid-metal interface (PLMI) during the experiment. The PLMI is prone to several forms of hydrodynamic and magnetohydrodynamic instabilities, including MRT instability, during the compression phase. The growth of instability at the PLMI disrupts implosion symmetry and introduces plasma contamination. 

Of particular interest, General Fusion Inc. is pioneering a novel implementation of the MTF concept. This approach involves the injection of plasma into the reactor core, which is subsequently compressed to achieve fusion conditions through the inward motion of a liquid-metal liner. The inward motion of the liquid liner can be accomplished either dynamically, by using mechanical pistons, or by utilizing magnetic forces~\citep{laberge2008, suponitsky2014}. One of the critical challenges within this approach pertains to maintaining the stability of the PLMI during the compression process. The occurrence of any instabilities at the PLMI has the potential to disrupt the plasma's purity, leading to plasma quenching. Throughout the compression process, the inner interface of the liquid metal is susceptible to RT instability. Numerous studies within the literature have studied the growth of RT instability at this inner interface, often exploring the implementation of liner rotation as a strategy to mitigate such instability~\citep{huneault2019rotational, avital2020}. Through rotation, a centrifugal force is generated, acting in opposition to the inward acceleration of the liner—an important mechanism for stabilization. 
% Another important feature is to maintain a smooth inward motion for the liquid liner since any perturbations emerging at the outer interface of the liner can propagate through the inner surface, i.e., the PLMI. 
On the other hand, during the magnetic compression of a liner, instabilities might arise from a combination of magnetic and hydrodynamic pressure. Consequently, a more comprehensive investigation is warranted to understand how perturbations on the MRT unstable interface would grow during the compression.

When a liquid liner of finite thickness is decelerated towards a low-density region such as a vacuum or plasma, one interface is subjected to MRT instability, while the other interface is stable. However, perturbations present on the unstable interface may feed through to the stable one. Additionally, the development of perturbations on the stable interface affects the temporal evolution of instabilities on the unstable interface throughout a complete cycle of a fusion reactor's operation. As a result, the feedthrough factor, which quantifies the effect of perturbation growth at one interface on the other interface, is introduced to quantify the impact of feedthrough for different scenarios. The MRT instability along with the feedthrough issue holds significant implications across different fusion approaches such as ICF, MTF, various Z-pinch configurations~\citep{chittenden2008}, and other liner-driven implosion applications. Among the earliest studies on the RT instability feedthrough one can refer to the study of Taylor~\cite{Taylor1950}. This effort was followed by other researchers, e.g., Mikaelian~\cite{mikaelian1985, mikaelian1990, mikaelian1995} studied RT and Richtmyer--Meshkov instabilities and the feedthrough effect in finite thickness fluid layers. 

The pioneering study of Harris~\cite{harris1962} on MRT feedthrough analytically examined the MRT instability of a collapsing cylindrical shell. This study was limited to cases where magnetic field lines remained unbent, resulting in a feedthrough factor identical to that reported by Taylor~\cite{Taylor1950} for the pure hydrodynamic case~\citep{lau2011anisotropy}. Subsequently, more comprehensive investigations were undertaken, accounting for factors such as magnetic tension and the anisotropic nature of MRT instability. Lau \textit{et al.}~\cite{lau2011anisotropy} conducted an analytical study on the MRT instability and feedthrough in a finite slab thickness using an ideal magnetohydrodynamic (MHD) model. Their examination allowed for the slab to experience acceleration resulting from the interplay of magnetic and fluid pressures. The Lau ~\textit{et al.}~\cite{lau2011anisotropy} study highlights the distinguishing characteristic of MRT instability—its anisotropic nature—setting it apart from the hydrodynamic RT instability.
% This study, thus, captured the distinguishing characteristic of MRT instability—its anisotropic nature—setting it apart from the hydrodynamic RT instability.

To elaborate further on the anisotropic nature of the MRT instability, one may consider a plasma slab supported by magnetic pressure in the presence of a downward gravitational force. In this scenario, plasma, a denser fluid, overlays the lighter medium of magnetic field lines; therefore, the interface is RT unstable. In instances where the magnetic field lines are orthogonal to initial perturbations of the interface, the growth rate of the instability aligns with that of the hydrodynamic case, $\left(k a\right)^{\frac{1}{2}}$. However, for situations where $\boldsymbol{k} \cdot \boldsymbol{B} \neq 0$, the MRT instability growth rate falls below $\left(k a\right)^{\frac{1}{2}}$ due to the influence of magnetic tension originating from the bent magnetic field lines~\citep{weis2015thesis}. This observation highlights the anisotropic behaviour intrinsic to MRT instability, setting it apart from classical RT instability. Subsequently, Weis \textit{et al.}~\cite{weis2014temporal} developed a theoretical expression to characterize the evolution of a surface ripple on a finite plasma slab which is MRT unstable over a limited time span. In their investigation, a finite plasma slab was confined between two perfect conductors and the obtained solution relied on the WKBJ approach. Each region may have an arbitrary magnetic field value with an arbitrary direction parallel to the interface. In their study, the ideal MHD model, along with linear theory, was employed, assuming all three regions to be incompressible with constant properties in each region. The general dispersion relation in the Cartesian coordinate system, a generalization of prior works~\citep{Taylor1950,harris1962,chandrasekhar1981hydrodynamic}, was derived along with the feedthrough factor. While the introduced model proved generally useful for studying instabilities in liner implosions, its accuracy is mainly limited when relatively large wavelength perturbations are present on the unstable interface. In cases where short wavelength perturbations exist and the resistivity of the regions cannot be neglected, the analytical solution is not quantitatively valid and can only be used for qualitative analysis~\citep{weis2015thesis}. Later on, this effort was extended to the cylindrical coordinate system, which is closer to the geometry of Z-pinch and allows the two well-known current carrying instabilities in cylindrical liners, i.e., sausage and kink modes, to appear~\citep{weis2015thesis}. For cylindrical geometries, Weis~\cite{weis2015thesis} also presented the analytical solution for the instability growth rate using the sharp boundary model in cylindrical coordinates under the ideal MHD assumption. The combined MRT and kink mode was reported as one of the main sources of instability in magnetized implosions in cylindrical geometries based on the obtained analytical results~\citep{weis2015thesis}. These findings were also verified by experimental and numerical studies~\citep{weis2014temporal, weis2015thesis}.

One of the main shortcomings in the aforementioned studies is the lack of consideration for the effect of liner resistivity. Generally, in cases where magnetic field lines and perturbation vectors are aligned, the MRT instability growth rate decreases due to the additional energy required to bend magnetic field lines according to the frozen-in law of ideal MHD cases. However, the existence of resistivity can reduce the stabilizing effect of the magnetic field~\citep{weis2014temporal}. This behaviour was observed in several studies focusing on the magnetic diffusion effect on the RT growth of the unstable interface of high-energy-density plasma in a constant background magnetic field in the whole domain~\citep{bera2022effect, samulski2022deceleration,barbeau2022design}. However, for our problem of interest, i.e., an initially magnetic-field-free liquid-metal liner undergoing MRT instability with the axial magnetic field present in the lower-density region, a more rigorous study is warranted to examine the effect of liner resistivity on perturbation growth and feedthrough.
% Therefore, a more rigorous study is needed to examine the effect of liner resistivity on perturbation growth and the feedthrough factor. 

The present study focuses on addressing this gap and aims to expand the existing knowledge of the impact of liner finite resistivity on perturbation growth. This will provide insight into how transitioning from the ideal MHD case to the resistive case affects the initial stages of MRT instability growth. The effect of surface tension on MRT growth is also investigated, since liquid-metal liners have relatively high surface tension. To this end, a novel level set-based two-phase incompressible MHD solver capable of examining both perfectly conductive and resistive fluids is introduced within the finite-difference framework. The detailed description of the problem investigated in this study is presented in Sec.~\ref{sec:Problem}, followed by the derivation of the dimensionless parameters describing the physics of the problem. The implementation of the numerical solver and the numerical setup employed for the simulations are discussed in Sec.~\ref{sec:numericalSolver}. The analytical solution of the problem under the ideal MHD assumption is given in Sec.~\ref{sec:anal}. Subsequently, the numerical results regarding the MRT instability growth and feedthrough in an initially magnetic-field-free liquid-metal slab in different regimes are reported in Sec.~\ref{sec:results}. Lastly, a detailed discussion of the results is provided in Sec.~\ref{sec:discussion}, and the study is summarized in Sec.~\ref{sec:conclusion}.

% \textcolor{red}{The detailed description of the problem investigated in this study is presented in Sec.~\ref{sec:problemDescription}, and the implementation of the numerical solver is discussed in Sec.~\ref{sec:numericalSolver} followed by the numerical setup employed for the simulations in Sec.~\ref{sec:sim_IC}. Lastly, this computational capability is utilized to study the MRT instability growth in an initially magnetic-field-free liquid-metal liner of finite thickness in the initial stages of compression. The effects of different Alfv\'en numbers, magnetic Reynolds numbers, and Bond numbers are studied on MRT instability growth and feedthrough, represented in Sec.~\ref{sec:results}. The study is concluded in Sec.~\ref{sec:conclusion}.}

\section{\label{sec:Problem} Problem description and formulation}
This section provides a detailed description of the problem under study, followed by the corresponding set of governing equations solved within the implemented numerical solver.

\subsection{\label{sec:problemDescription} Problem of interest}
\begin{figure}
\centering
\includegraphics[width=0.7\textwidth]{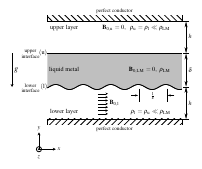}
\caption{\label{fig:ProblemDescription} The schematic of the problem of interest. The MRT growth of the initially seeded perturbation with the wavenumber denoted by $k$ at the lower interface and its impact on the upper interface are examined.}
\end{figure}

The schematic of the studied problem is depicted in Fig.~\ref{fig:ProblemDescription}, where a liquid-metal liner of thickness $\delta$ is bounded by two lighter regions, referred to as lower and upper layers. The lower and upper layers are assumed to have the same material properties such as density and electrical conductivity. Additionally, for simplification, the thickness of these two layers is assumed to be similar, denoted by $h$.

For the fusion applications described in Sec.~\ref{sec:introduction}, the MRT instability arises from the acceleration of the liner bounded by a lower density region, with the acceleration of the lower density region toward the greater density region being unstable. For the purposes of the present study, we will consider an analogous problem with the analysis being performed in the reference frame of the liner. Rather than accelerating the liner, a body force term denoted by $g$ will be introduced that has a similar effect on liner dynamics. In this reference frame, the body force term will be directed \emph{downwards}, from the heavier fluid (i.e., the liquid liner) toward the lighter one (i.e., the lower air layer), a scenario that is RT unstable, as shown in Fig.~\ref{fig:ProblemDescription}. The upper interface ideally remains stable; however, perturbations on the lower interface can affect the upper surface (i.e., feedthrough effect). Our interest lies in examining the RT instability growth and feedthrough at the lower and upper interfaces, respectively. Each region is assumed to be incompressible, with the lower and upper layers bounded by perfect conductors. Initially, no magnetic field is present in the liquid-metal liner and the upper layer, while a constant horizontal magnetic field, $B_\mathrm{0,l}$, is imposed in the lower region.

In the course of the numerical campaign conducted in this study, the effect of magnetic field strength present in the lower region on MRT growth and feedthrough is investigated for different liner thicknesses and perturbation wavenumbers. This analysis is then repeated for the case of a resistive liquid-metal liner. For a more rigorous examination and to facilitate the application of this analysis across various scenarios, a set of non-dimensional equations is solved, presented in the following section, and the influence of governing dimensionless numbers on instability growth is explored.

\subsection{\label{sec:MHDEqs} Governing equations }
To better characterize the effect of different parameters such as magnetic tension, magnetic diffusion, and surface tension on the MRT instability growth, the dimensionless form of the governing equations is presented. The reference values used to non-dimensionalize the equations are as follows
\begin{eqnarray}
\boldsymbol{x}^*=\frac{\boldsymbol{x}}{L_\mathrm{ref}}=\frac{\boldsymbol{x}}{\delta}, \         \
\boldsymbol{u}^*=\frac{\boldsymbol{u}}{U_\mathrm{ref}}=\frac{\boldsymbol{u}}{\sqrt{\delta g}}, \        \ t^*= \frac{t}{T_\mathrm{ref}}= t\sqrt{\frac{g}{\delta}}, \\ \nonumber
\          \ \boldsymbol{B}^*=\frac{\boldsymbol{B}}{B_\mathrm{ref}}=\frac{\boldsymbol{B}}{B_{0,\mathrm{l}}}, \      \ \text{and}\          \ p^*=\frac{p}{\rho_\mathrm{l} U_\mathrm{ref}^2},
\end{eqnarray}
where the superscript ``$*$" represents the dimensionless variables. The thickness of the liner, the gravitational acceleration, and the initial uniform axial magnetic field at the lower region are denoted by $\delta$, $g$, and $B_{0,\mathrm{l}}$, respectively. Variables $\boldsymbol{u}$ and $p$ are the velocity and pressure fields, respectively, and $\rho_\mathrm{l}$ is the density of the liquid-metal liner. \\
Consequently, the non-dimensional form of the two-phase incompressible resistive MHD equations for inviscid flows is written as
\begin{subequations}
\begin{eqnarray}
\label{eq:mom}
\frac{\partial \boldsymbol{u}^*}{\partial t^*} + \nabla \cdot \left(\boldsymbol{u}^* \boldsymbol{u}^* \right) = -\frac{1}{\rho} \nabla p^* + \frac{\boldsymbol{g}}{g} + 
\frac{1}{\rho} \, \frac{1}{\mathrm{Bo}} \, \boldsymbol{F}_\mathrm{ST} + \frac{1}{\rho} \, \frac{1}{\mathrm{Al}^2} \, \nabla \cdot \tau^\mathrm{M},
\end{eqnarray}
\begin{equation}
    \label{eq:vel-div}
    \nabla \cdot \boldsymbol{u}^* = 0,
\end{equation}
\begin{eqnarray}
    \label{eq:induction}
    \frac{\partial \boldsymbol{B}^*}{\partial t^*} = \nabla \cdot \left(\boldsymbol{B}^* \boldsymbol{u}^* - \boldsymbol{u}^* \boldsymbol{B}^* \right) + \frac{1}{\mathrm{Re}_\mathrm{m}} \nabla \cdot \left(\frac{1}{\sigma_\mathrm{e}} \nabla \boldsymbol{B}^* \right),
\end{eqnarray}
\begin{equation}
    \label{eq:mag-div}
    \nabla \cdot \boldsymbol{B}^* = 0,
\end{equation}
\end{subequations}
where the three dimensionless parameters, Bond number, $\mathrm{Bo}$, Alfv\'en number, $\mathrm{Al}$, and magnetic Reynolds number, $\mathrm{Re}_\mathrm{m}$, are defined as

\begin{equation}
     \mathrm{Bo} = \frac{\rho_\mathrm{l} \, L_\mathrm{ref} U_\mathrm{ref}^2}{\sigma}, \         \ \mathrm{Al} = \frac{U_\mathrm{ref}}{U_\text{Alfv\'en}}, \ \  \text{and} \    \  \mathrm{Re}_\mathrm{m} = \frac{ L_\mathrm{ref} \, U_\mathrm{ref}}{\lambda_\mathrm{m,l}}.
\end{equation}
 Variables $\rho$ and $\sigma_\mathrm{e}$ are the dimensionless density and electrical conductivity, respectively, given as 
\begin{equation}
    \rho = \rho_\mathrm{r} + \left(1 - \rho_\mathrm{r} \right)\psi \     \  \  \ \text{and} \    \ \      \ \sigma_\mathrm{e} = \sigma_\mathrm{e, r} + \left(1 - \sigma_\mathrm{e, r} \right)\psi,
\end{equation}
where $\rho_\mathrm{r} = \rho_\mathrm{g}/\rho_\mathrm{l}$ and $\sigma_\mathrm{e, r} = \sigma_\mathrm{e, g}/\sigma_\mathrm{e, l}$, with subscripts ``$\mathrm{l}$'' and ``$\mathrm{g}$'' representing the liquid and gas phases, respectively. The scalar variable $\psi$ is the level set function utilized to capture the interface between the liquid and gas phases, varying between $0$ (gas phase) to $1$ (liquid phase). More detail regarding the implemented level set method is given in the next section.

The Bond number quantifies the intensity of the surface tension force, $\boldsymbol{F}_\mathrm{ST}$, with variable $\sigma$ denoting the surface tension coefficient of the liquid liner. In order to account for the effect of magnetic forces on the fluid motion, the Lorentz force is incorporated into the momentum equation, Eq.~(\ref{eq:mom}). The Lorentz force, which quantifies the force experienced by conducting fluids due to electromagnetic fields, is given as $\boldsymbol{J} \times \boldsymbol{B}$, where $\boldsymbol{J}$ is the electric current density. This force can be written in the form of the Maxwell stress tensor, $\tau^{\mathrm{M}}$, given as \citep{davidson2002}
\begin{equation}
    \nabla \cdot \tau_{ij}^{\mathrm{M}} = \left({B}^*_i {B}^*_j - \frac{|{B}^*|^2}{2} \delta_{ij} \right),
\end{equation}
which is the conservative representation of this force employed for numerical discretization, as explained in the subsequent section.\\
The dimensionless Alfv\'en number is the ratio between the characteristic velocity to the Alfv\'en wave speed which is defined as 
\begin{equation}
    U_\text{Alfv\'en} = \frac{B_0}{\sqrt{\mu_\mathrm{m} \, \rho_\mathrm{l}}},
\end{equation}
where $\mu_\mathrm{m}$ stands for the magnetic permeability of the corresponding medium. The Alfv\'en number indicates the relative strengths of the magnetic and inertial forces present in MHD flows. A high Alfv\'en number implies that inertial forces dominate over the existing magnetic forces, while reducing the Alfv\'en number results in regimes where magnetic forces are more pronounced.

Equation~(\ref{eq:induction}), known as the induction equation, represents the evolution of the magnetic field due to the advection, $\nabla \cdot \left(\boldsymbol{B}^*\boldsymbol{u}^* - \boldsymbol{u}^*\boldsymbol{B}^* \right)$, and diffusion, $\nabla \cdot  \left( \frac{1}{\sigma_{\mathrm{e}}} \nabla \boldsymbol{B}^* \right)$. The induction equation comes from combining the Faraday, Maxwell--Amp\`ere, and Ohm's laws, and the magnetic Reynolds number quantifies the relative strengths of advection and diffusion of a magnetic field, where $\lambda_\mathrm{m,l} = 1/\left(\mu_\mathrm{m}\, \sigma_\mathrm{e,l} \right)$ illustrates the magnetic diffusivity of the liquid liner. 

It is noteworthy to mention that momentum and induction equations, Eqs.~(\ref{eq:mom}) and (\ref{eq:induction}), should be solved while satisfying the divergence-free constraint for both velocity and magnetic fields, Eqs.~(\ref{eq:vel-div}) and (\ref{eq:mag-div}). The following section describes the implemented numerical toolkit to solve Eqs.~(\ref{eq:mom}-\ref{eq:mag-div}) in more detail.

\section{\label{sec:numericalSolver} Implementation of a two-phase MHD numerical solver }
This section presents the methodology employed for simulating two-phase MHD flows, along with a description of the initial conditions and boundary conditions imposed in the simulations.

% In this section, the governing set of equations is presented, followed by an introduction to the implemented solver for two-phase incompressible ideal/resistive MHD flows.

\subsection{\label{sec:discretization} Grid arrangement and discretization}
\begin{figure}
\centering
\includegraphics[width=0.8\textwidth]{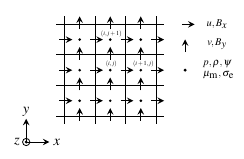}
\caption{\label{fig:grid} Staggered grid system in Cartesian coordinates. In the staggered grid arrangement, the values of scalar fields such as pressure ($p$), level set function ($\psi$), density ($\rho$), magnetic permeability ($\mu_\mathrm{m}$), and electrical conductivity ($\sigma_\mathrm{e}$) are defined at cell centers. Velocity components, $u$ and $v$, along with magnetic field components, $B_x$ and $B_y$, are defined at cell faces.}
\end{figure}

The staggered grid arrangement is used in the implemented solver, shown in Fig.~\ref{fig:grid}. In this computational grid system, velocity and magnetic field values are represented at cell faces, $(i+\frac{1}{2}, j)$ and $(i, j+\frac{1}{2})$, while other scalar variables such as pressure, level set, and properties such as density and electrical conductivity are defined at cell centers, $(i, j)$.

The implemented two-phase resistive MHD solver is an extension to the work of Makaremi-Esfarjani \textit{et al.}~\cite{makaremi}, where a detailed description of a two-phase incompressible solver for magnetic flows is provided. In that study, a fifth-order conservative level set method was employed to capture the evolution of the interface, coupled with the projection-based incompressible solver. The solver demonstrated second-order accuracy and showed excellent performance in handling high density ratios across the interface. Additionally, the surface tension force was modelled using the continuum interface force (CSF) approach. In this section, we focus on extending the mentioned solver to the resistive MHD case. Therefore, the discretization of the magnetic forces along with the implementation of the induction equation while satisfying the divergence-free condition are discussed. Interested readers may refer to Makaremi-Esfarjani \textit{et al.}~\cite{makaremi} for more details regarding the implementation of the two-phase solver for hydrodynamic and magnetic flows.

Using the notation introduced in previous studies for the second-order finite-difference and second-order interpolation operators~\citep{morinishi1998fully, desjardins2008high, makaremi}, the discretization of the $x-$component of the Lorentz force in the two-dimensional computational domain, $(x, y)$, with velocity and magnetic fields denoted by $(u, v)$ and  $(B_x, B_y)$, respectively, is given as
\begin{eqnarray}
    F_{\mathrm{Lorentz}, x} \Bigg|_ {i+\frac{1}{2},j} = \frac{1}{\rho} \frac{1}{\mathrm{Al}^2} \left[ \frac{\partial B_x B_x}{\partial x} + \frac{\partial B_x B_y}{\partial y} - \frac{1}{2} \left( \frac{\partial B_x^2}{\partial x} + \frac{\partial B_y^2}{\partial x} \right) \right] \\ \nonumber
    = \frac{1}{\overline{\rho}^{2\mathrm{nd} \, x}} \frac{1}{\mathrm{Al}^2} \left[ \frac{\delta_2 \, \left(\overline{B_x}^{2\mathrm{nd} \, x} \, \overline{B_x}^{2\mathrm{nd} \, x} \right) } {\delta_2 \, x}  + 
    \frac{\delta_2 \, \left(\overline{B_x}^{2\mathrm{nd} \, y} \, \overline{B_y}^{2\mathrm{nd} \, x} \right) } {\delta_2 \, y} - \frac{1}{2} \left(\overline{\frac{\delta_2 \, B_x^2}{\delta_2 x}}^{2\mathrm{nd} \, x} + \overline{\frac{\delta_2 \, B_y^2}{\delta_2 \, x}}^{2\mathrm{nd} \, y} \right) \right].
\end{eqnarray}
The term $F_{\mathrm{Lorentz}, x}$ signifies the Lorentz force present in the $x-$component of the momentum equation. Consequently, these values need to be computed at cell faces along the $x-$direction, $(i+\frac{1}{2}, j)$. Similarly, the $y-$component of the Lorentz force is determined at cell faces $(i,j+\frac{1}{2})$. 

The discretization of the induction equation for $B_x$ component, while considering the electrical conductivity jump across the interface, is given as 
\begin{eqnarray}
     \frac{\partial B_x}{\partial t} \Bigg|_ {i+\frac{1}{2},j}=\frac{\partial}{\partial y}\left(u B_y - B_x v \right) + \frac{\partial}{\partial x} \left( \frac{1}{\sigma_\mathrm{e}} \frac{\partial B_x}{\partial x} \right) + \frac{\partial}{\partial y} \left( \frac{1}{\sigma_\mathrm{e}} \frac{\partial B_x}{\partial y}\right) \\ \nonumber
      =\frac{\delta_2 \, \left( \overline{u}^{2\mathrm{nd} \, y} \, \, \overline{B_y}^{2\mathrm{nd} \, x} \right)}{\delta_2 \, y} - \frac{\delta_2 \, \left(\overline{B_x}^{2\mathrm{nd} \, y} \, \,  \overline{v}^{2\mathrm{nd} \, x} \right)}{\delta_2 \, y}
       + \frac{\delta_2 \, \left(\frac{1}{\sigma_\mathrm{e}} \, \frac{\delta_2 \, B_x}{\delta_2 \, x} \right)}{\delta_2 \, x} + \frac{\delta_2 \, \left( \frac{1}{\overline{\overline{\sigma_\mathrm{e}}^{2\mathrm{nd} \, x}}^{2\mathrm{nd} \, y}} \, \frac{\delta_2 B_x}{\delta_2 \, y} \right)}{\delta_2 \, y} ,
\end{eqnarray}
at cell faces $(i+\frac{1}{2}, j)$. The electrical conductivity value, $\sigma_\mathrm{e}$, should also be updated alongside other properties such as density, according to the updated location of the interface determined by the level set function at each time step.
In a similar fashion, the $B_y$ component discretization is derived at cell faces across the $y-$direction, $(i, j+\frac{1}{2})$. 

The solution of the induction equation, Eq.~(\ref{eq:induction}), does not necessarily satisfy the magnetic field divergence-free constraint. Maintaining the divergence-free condition of the magnetic field significantly influences the accuracy of the numerical solver, and violating this constraint results in unphysical numerical solutions. Various schemes, such as the eight-wave formulation, constrained transport, and projection scheme have been proposed in the literature to impose the divergence-free condition of the magnetic field~\citep{toth2000b}. In this study, we have employed the projection scheme, where the computed magnetic field from the induction equation is projected into a divergence-free field~\citep{brackbill1980effect}. The calculated magnetic field from Eq.~(\ref{eq:induction}) at time step $n+1$, denoted by $\tilde{\boldsymbol{B}}^{n+1}$, can be written as a summation of a curl and a gradient as $\tilde{\boldsymbol{B}}^{n+1} = \nabla \times \boldsymbol{A}+\nabla \phi$, where only the curl of vector potential $\boldsymbol{A}$ represents the physical part of the magnetic field solution. Taking the divergence of the mentioned equation results in the following Poisson equation for the scalar field $\phi$ 
\begin{equation}
    \nabla \tilde{\boldsymbol{B}}^{n+1} = \nabla^2 \phi.
\end{equation}
Solving this Poisson equation and finding the scalar field $\phi$, the magnetic field $\tilde{\boldsymbol{B}}^{n+1}$ will be projected into the divergence-free field as 
\begin{equation}
    \boldsymbol{B}^{n+1} = \tilde{\boldsymbol{B}}^{n+1} - \nabla \phi,
\end{equation}
where $\boldsymbol{B}^{n+1}$ is the magnetic field solution at time step $n+1$ which satisfies the divergence-free constraint.

The third-order explicit Runge--Kutta method is used for the temporal integration~\citep{gottlieb1998total}, and based on the CFL condition, the stability constraint for the time step due to convection, diffusion, and surface tension terms is given as
\begin{equation}
\label{eq:cfl}
    \Delta t \leq \mathrm{min} \left(\frac{\Delta x}{\mathrm{max}(||\mathbf{u}|| + ||\mathbf{u}_\text{Alfv\'en}||)}, \frac{1}{4} \frac{\Delta x^2}{\mathrm{max}(\lambda_\mathrm{m,l}, \lambda_\mathrm{m,g})}, \sqrt{\frac{{\Delta x}^3 \left(\rho_\mathrm{l} + \rho_\mathrm{g} \right)}{4 \pi \sigma}} \right).
\end{equation}
As can be observed in Eq.~(\ref{eq:cfl}), the velocity of the Alfv\'en wave and magnetic diffusivity should also be considered in determining the time step compared to the hydrodynamic case. Consequently, for high magnetic field values or low magnetic Reynolds numbers, the system of governing equations becomes stiff, resulting in excessively small time steps. Due to this numerical stiffness, different semi-implicit and implicit schemes for resistive MHD equations have been proposed in the literature to address the stiffness issue associated with Alfv\'en waves, such as the implicit solver introduced for reduced resistive MHD equations by Chac\'on~\cite{chacon2002implicit}. However, for our problem of interest, the explicit Runge--Kutta scheme was deemed to be sufficient and straightforward to implement.

In Appendix~\hyperlink{appA2}{A}, the accuracy, convergence, and performance of the introduced two-phase incompressible MHD solver are examined. Interested readers can refer to this section for more details.

\subsection{\label{sec:sim_IC} Numerical simulation setup}
The description of the numerical simulation setup and boundary conditions is discussed herein. In this study, the MRT instability of a planar liquid liner slab in a two-dimensional domain is investigated. Although in most cases three-dimensional simulations are needed to fully capture instability growth, two-dimensional studies still provide valuable insights. Furthermore, since our primary focus lies on the initial stages of perturbation growth parallel to the initially imposed axial magnetic field, the perturbation growth mainly occurs within the $x-y$ plane; and, as a result, the two-dimensional assumption is reasonable.
% Furthermore, in cases where the perturbation growth rate in two directions is considerably smaller compared to the third direction, the two-dimensional approximation remains valid \citep{bera2022effect}. Since our primary focus lies on the initial stage of perturbation growth parallel to the initially imposed transverse magnetic field, the perturbation growth occurs within the $x-y$ plane; and, as a result, the two-dimensional assumption is reasonable.

In the initial condition, an inviscid liquid liner with the density and thickness of $\rho_\mathrm{l} = 500 \, \mathrm{kg}/\mathrm{m}^3$ and $\delta=\pi/6$, respectively, is surrounded by air in the computational domain of $ (x, y) \in [0, \pi] \times [0, \pi]$. 
 The density ratio is set to $\rho_\mathrm{r}=0.002$ for all the presented simulations. Therefore, the Atwood number is almost 1, which is defined as $\mathrm{At}=(\rho_\mathrm{l} - \rho_{\mathrm{g}})/(\rho_\mathrm{l} + \rho_\mathrm{g})$. The lower interface of the liquid liner is initially perturbed with a sinusoidal perturbation with the amplitude of $\pi/40$, and a uniform magnetic field $B_{0,\mathrm{l}}$ is imposed in the lower region consisting of air. All simulations are performed with a constant CFL number of 0.1.

A periodic boundary condition is set along the $x-$direction, and the top and bottom boundaries are assumed to be perfectly conducting rigid walls. Therefore, the slip boundary condition is used for the velocity field at the top and bottom boundaries. To impose the perfectly conducting boundary condition for the magnetic field, the normal component of the magnetic field at the walls is set to zero.

To ensure the results of this study are relevant to the various fusion approaches discussed in the introduction, the parameter space for the three key non-dimensional numbers is presented in Table~\ref{tab:1} for two fusion approaches utilizing a metal liner for compression: MTF and MagLIF. For the MTF approach, the characteristic properties of the General Fusion Inc. power plant design are used, and the working liquid metal is assumed to be liquid lithium. The properties reported in Table~\ref{tab:1} for MagLIF are based on the Sandia National Laboratory (SNL) Z facility, where a beryllium liner is utilized to compress the deuterium-tritium (D-T) fuel. 

Table~\ref{tab:1} offers insight into the approximate values of the three dimensionless parameters encountered in practical scenarios. In this study, we mainly focus on time scales during which the RT instability remains in laminar regimes and smaller-scale structures have yet to emerge. Although over the entire liner compression process the incompressibility assumption may no longer hold, studying MRT instability growth in the initial stages of compression, where the liquid liner can be treated as incompressible, is of primary interest. The parameters represented in Table~\ref{tab:1} are helpful in guiding our focus on regimes comparable to those in fusion reactors. Considering the capabilities and limitations of the implemented solver, the conducted test cases are designed so that the corresponding parameters fall within the parameter space introduced here, as presented in the result section.
\begin{table*}[]
% \scriptsize
\centering
\caption{The parameter space of the Bond number, Alfv\'en number, and magnetic Reynolds number for two fusion approaches: MTF and MagLIF.}
\label{tab:1}
\begin{tabular}{cccc}
\hline
& & MTF$^{\hyperlink{1}{b}}$ & MagLIF$^{\hyperlink{2}{c}}$ \\
\hline
& $\rho_{\mathrm{l}}$ [kg m\textsuperscript{-3}] & 500 & 1850 \\
& $\sigma$ [N m\textsuperscript{-1}] & 0.4 & 1.5$^{\hyperlink{1}{d}}$ \\
 & $\sigma_\mathrm{e}$ [S m\textsuperscript{-1}] & $4 \times 10^{6}$ & $2.5 \times 10^{7}$ \\
properties$^{\hyperlink{1}{a}}$ & $\mu_{\mathrm{m}}$ [H m\textsuperscript{-1}] & $4\pi \times 10^{-7}$ & $4\pi \times 10^{-7}$ \\
& $\delta$ [m] & 0.4 & $10^{-4}$ \\ & \vspace*{7mm}
\multirow{2}{*}
 {\shortstack{$B_0$ [T]}} & \multirow{2}{*} {\shortstack{0.7 (Uncompressed)\\70 (Compressed)}} & \multirow{2}{*} {\shortstack{10 (Uncompressed)\\100 (Compressed)}} \\ \vspace*{7mm}
& $g$ [m s\textsuperscript{-2}] & $10^{5}$ & $10^{11}$ \\
& $\mathrm{Bo}$ & $\approx 10^{7}$ & $\approx 10^{6}$ \\
parameters& $\mathrm{Al}$ & $0.07-7$ & $1.5-15$ \\
& $\mathrm{Re}_\mathrm{m}$ & 400 & 10 \\

\hline
\end{tabular}

\hypertarget{1}{\textcolor{red}{$^a$} The presented values are approximations, as the properties of liquid metal undergo significant changes at high temperatures, and in most cases, exact values are not known.}

\hypertarget{1}{\textcolor{red}{$^b$} The liquid lithium properties are taken from Davison~\cite{davison1968compilation}}.

\hypertarget{2}{\textcolor{red}{$^c$} The analytical expressions of beryllium properties are given by Tolias~\cite{tolias2022analytical}}.

\hypertarget{2}{\textcolor{red}{$^d$} Taken from Kumikov~\cite{kumikov1983measurement}}.
\end{table*}

\section{\label{sec:anal} Analytical solution}
In the present study, the numerical toolkit developed in Sec.~\ref{sec:numericalSolver} is employed to study the effect of magnetic tension and diffusion on the MRT instability growth. However, the analytical solution obtained by Weis \textit{et al.}~\cite{weis2014temporal} can be used to predict the MRT instability growth under the ideal MHD assumption within the linear regime and be compared to the numerical results. In their study, Weis \textit{et al.}~\cite{weis2014temporal} derived the analytical dispersion relation for the MRT instability growth of a finite plasma slab confined by two incompressible regions, assuming all regions to be perfectly conductive. This dispersion relation is reformulated for the problem described in Sec.~\ref{sec:problemDescription} based on the defined reference values. The resulting non-dimensional dispersion relation is as follows   
\begin{equation}
\label{eq:dispersion}
   A^* {\omega^*}^4 - B^* {\omega^*}^2 + C^*= 0,     
\end{equation}
where $\omega^*$ is the dimensionless growth rate, $\omega^*=\omega/\sqrt{k \, g}$, and coefficients $A^*$, $B^*$, and $C^*$ are given as 
\begin{subequations}
\begin{equation}
    A^* = 1 + \rho_{\mathrm{r}}^2 \, \coth(k^* h^*) \, \coth(k^* h^*) + 2 \, \rho_\mathrm{r} \, \coth(k^*  h^*) \, \coth(k^*),
\end{equation}
\begin{equation}
    B^* = \frac{k^*}{\mathrm{Al}^2} \, \coth(k^* h^*) \coth(k^*) + \frac{\rho_\mathrm{r} \, k^*}{\mathrm{Al}^2} \, \coth(k^* h^*) \, \coth(k^* h^*) ,
\end{equation}
and
\begin{equation}
    C^* = \left(1 - \rho_\mathrm{r} \right) \left(\rho_\mathrm{r} +  \frac{k^*}{\mathrm{Al}^2} \coth(k^* h^*) - 1 \right).
\end{equation}
\end{subequations}
Variables $k^*$ and $h^*$ are dimensionless wavenumber and lower/upper layer thickness, respectively, and are calculated as $k^* = k \, \delta$ and $h^* = h/ \delta$. The solution of the dispersion equation, Eq.~(\ref{eq:dispersion}), gives
\begin{equation}
   \label{eq:w2Solution}
    {\omega^*}^2 = \frac{1}{2A^*} \left(B^* \pm \sqrt{{B^*}^2 - 4 A^* C^*} \right).
\end{equation}
According to the energy principle for the MRT instability in the ideal MHD case, ${\omega^*}^2$ is always real; therefore, ${B^*}^2 - 4 A^* C^* \geq 0$~\citep{boyd2003physics}. Owing to this property, the solution of Eq.~(\ref{eq:w2Solution}) results in four $\omega^*$ values which are either purely real or imaginary and are in negative and positive pairs. Since instability occurs when ${\omega^*}^2<0$, the imaginary solution of $\omega^*$ with the greater value corresponds to the MRT instability growth rate of the most unstable mode.

\begin{figure}[]
% \hspace*{-1cm}
\centering
\subfloat[]{\includegraphics[width = 0.5\textwidth]{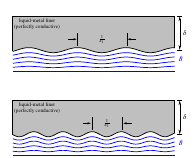}}\\
\hspace*{-1.8cm}
\subfloat[]{\includegraphics[width = 1\textwidth]{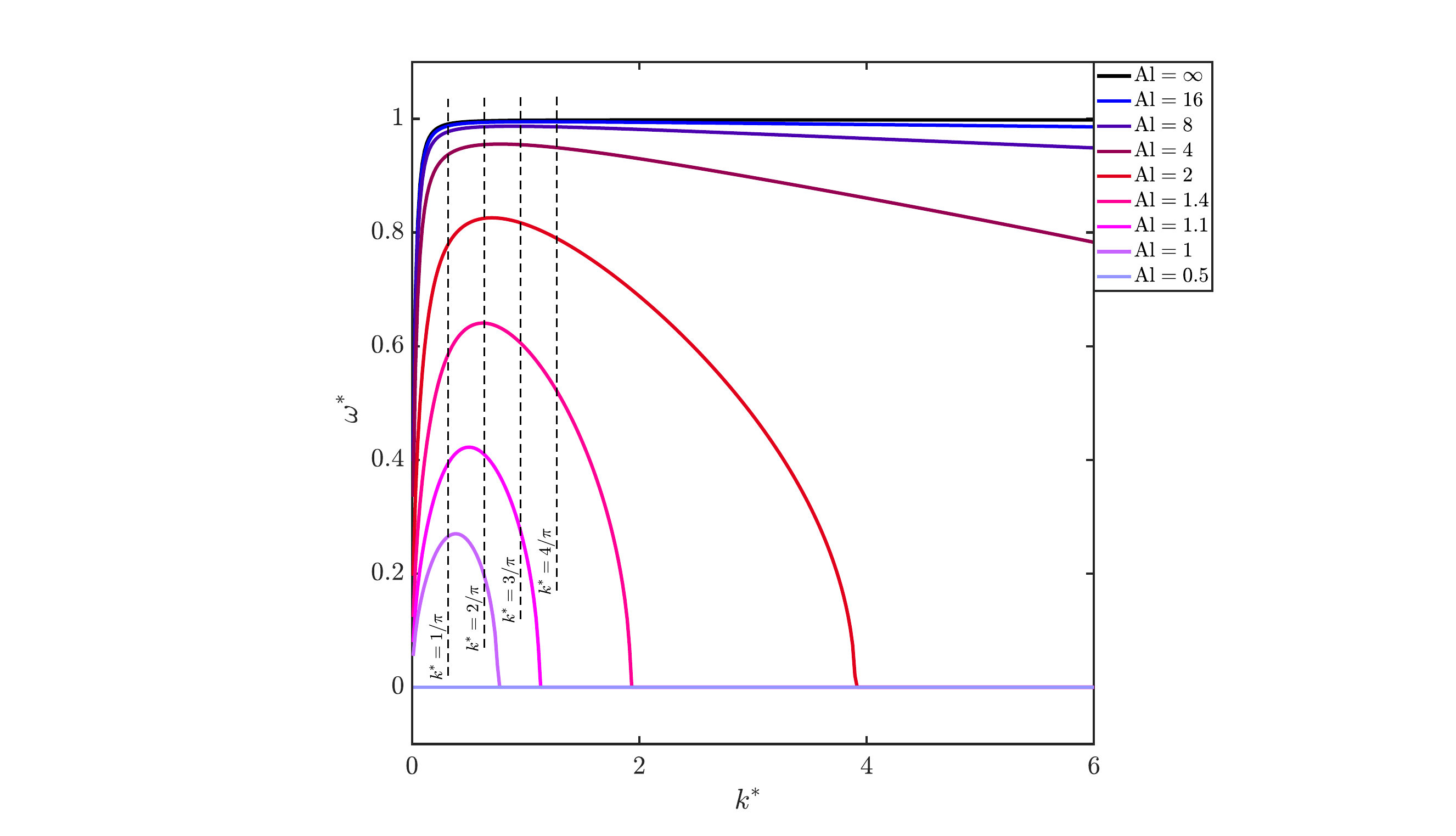}}
% \hspace*{1cm}
\caption{ (\textit{a}) Schematic of the problem for two different wavenumbers ($k_2>k_1$). Higher wavenumbers (i.e., shorter wavelengths) result in bent magnetic field lines with greater curvature. (\textit{b}) The non-dimensional MRT growth rate, $\omega^*$, as a function of dimensionless wavenumber, $k^*$, for nine different Alfv\'en numbers of $\infty$, $16$, $8$, $4$, $2$, $1.4$, $1.1$, $1$, and $0.5$.}
\label{fig:idealMHD1}
\end{figure}

% For the initial condition described in Sec.~\ref{sec:sim_IC} and nine different Alfv\'en numbers of $\infty$, $16$, $8$, $4$, $2$, $1.4$, $1.1$, $1$, and $0.5$, the MRT growth rate, $\omega^*$, is calculated for different wavenumbers, shown in Fig.~\ref{fig:idealMHD1}, using the derived analytical solution. 
For the initial condition described in Sec.~\ref{sec:sim_IC} and nine different Alfv\'en numbers of $\infty$, $16$, $8$, $4$, $2$, $1.4$, $1.1$, $1$, and $0.5$, the MRT growth rate, $\omega^*$, is calculated for different wavenumbers. The schematic of the problem for two different wavenumbers and the corresponding magnetic field configurations are depicted in Fig.~\ref{fig:idealMHD1}(a), while the corresponding analytical results are presented in Fig.~\ref{fig:idealMHD1}(b).
It is appreciated that generally, decreasing the Alfv\'en number, i.e., increasing the initial magnetic field present in the lower layer, $B_{0,\mathrm{l}}$, results in a reduced MRT growth rate. As shown in Fig.~\ref{fig:idealMHD1}(b), for $\mathrm{Al}=16$, the calculated MRT growth rate is slightly smaller compared to the case of pure hydrodynamic, $\mathrm{Al}=\infty$, and this difference becomes more pronounced for higher $k^*$ values.  This behaviour is consistent as the Alfv\'en number is decreased to 8 and below. As the magnetic field is increased, a critical magnetic field strength is reached beyond which the MRT instability is fully stabilized, attributed to the presence of an axial magnetic field. For example, for $\mathrm{Al}=0.5$ in the studied problem, the calculated MRT growth rate is zero (see Fig.~\ref{fig:idealMHD1}b). 
% The reduction in MRT growth with an increase in the imposed magnetic field can be justified using the frozen-in law, which states that in ideal MHD flows, magnetic field lines are attached to the velocity field. Therefore, as the instability ripples begin to grow, magnetic field lines trapped in the lower layer also start to ripple and bend. Consequently, the bent magnetic field lines experience tension, requiring additional energy to bend them, resulting in a decrease in instability growth rate.  

Another noteworthy observation from Fig.~\ref{fig:idealMHD1}(b) is the presence of a critical wavenumber, $k^*_{\mathrm{critical}}$, at which the MRT instability growth reaches its maximum value for a constant Alfv\'en number. For instance, according to Fig.~\ref{fig:idealMHD1}(b), for Alfv\'en numbers of 2 and 1.4, the critical wavenumber is close to $k^*=2/\pi$, corresponding to the peak MRT growth. This critical wavenumber decreases as the Alfv\'en number decreases, approaching zero for fully stable cases. Figure \ref{fig:idealMHD1}(b) further shows that, in general, the magnetic field exhibits a more stabilizing effect for higher wavenumbers, that is, shorter wavelengths. 

The MRT growth rate and feedthrough are numerically investigated for the four wavenumbers indicated in Fig.~\ref{fig:idealMHD1}(b), $k^*=1/\pi$, $2/\pi$, $3/\pi$, and $4/\pi$. The results from the numerical simulations for the ideal case are compared with the analytical ones in the following section.

% One may ascribe this behaviour to the induced tension in the magnetic field during the MRT growth. In the case of a single interface MRT instability, the bending of the magnetic field generates Alfv\'en waves, and the magnetic tension part of the Lorentz force is proportional to $\left(\boldsymbol{k} \cdot \boldsymbol{v}_\text{Al} \right)^2$. Hence, as the wavenumber increases, the stabilizing effect of the magnetic field becomes more pronounced, leading to further suppression of instabilities at shorter wavelengths.

\section{\label{sec:results} Results}
In this section, first, the effect of the Alfv\'en number on the MRT growth rate and feedthrough effect is studied for the ideal case, where all regions are assumed to be perfectly conductive. This effort is then extended to the resistive MHD case, examining the effect of liner finite resistivity on instability growth for different magnetic Reynolds numbers. Additionally, the effect of electrical conductivity jump across the interface on the MRT growth rate is investigated for different wavenumbers. Finally, the effect of liquid-metal surface tension on MRT growth for both ideal and resistive cases is presented.

% OLD version
% In this section, first, the effect of magnetic field strength on MRT growth is studied for the ideal case. The numerical results are compared to the analytical ones for different wavenumbers and Alfv\'en numbers. This effort is then extended to the resistive MHD case, examining the effect of liner finite resistivity on instability growth and feedthrough on the upper interface for different magnetic Reynolds numbers. Additionally, the effect of electrical conductivity jump across the interface on the growth rate is investigated for different wavenumbers. Finally, the effect of liquid-metal surface tension on MRT growth for both ideal and resistive cases is presented.

\subsection{\label{sec:idealMHD} Ideal MHD case }
\begin{figure}[]
\centering
% \hspace*{3.5cm}
\subfloat[]{\includegraphics[width = 0.95\textwidth]{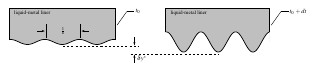}}\\
\hspace*{-2.4cm}
\subfloat[]{\includegraphics[width = 1.2\textwidth]{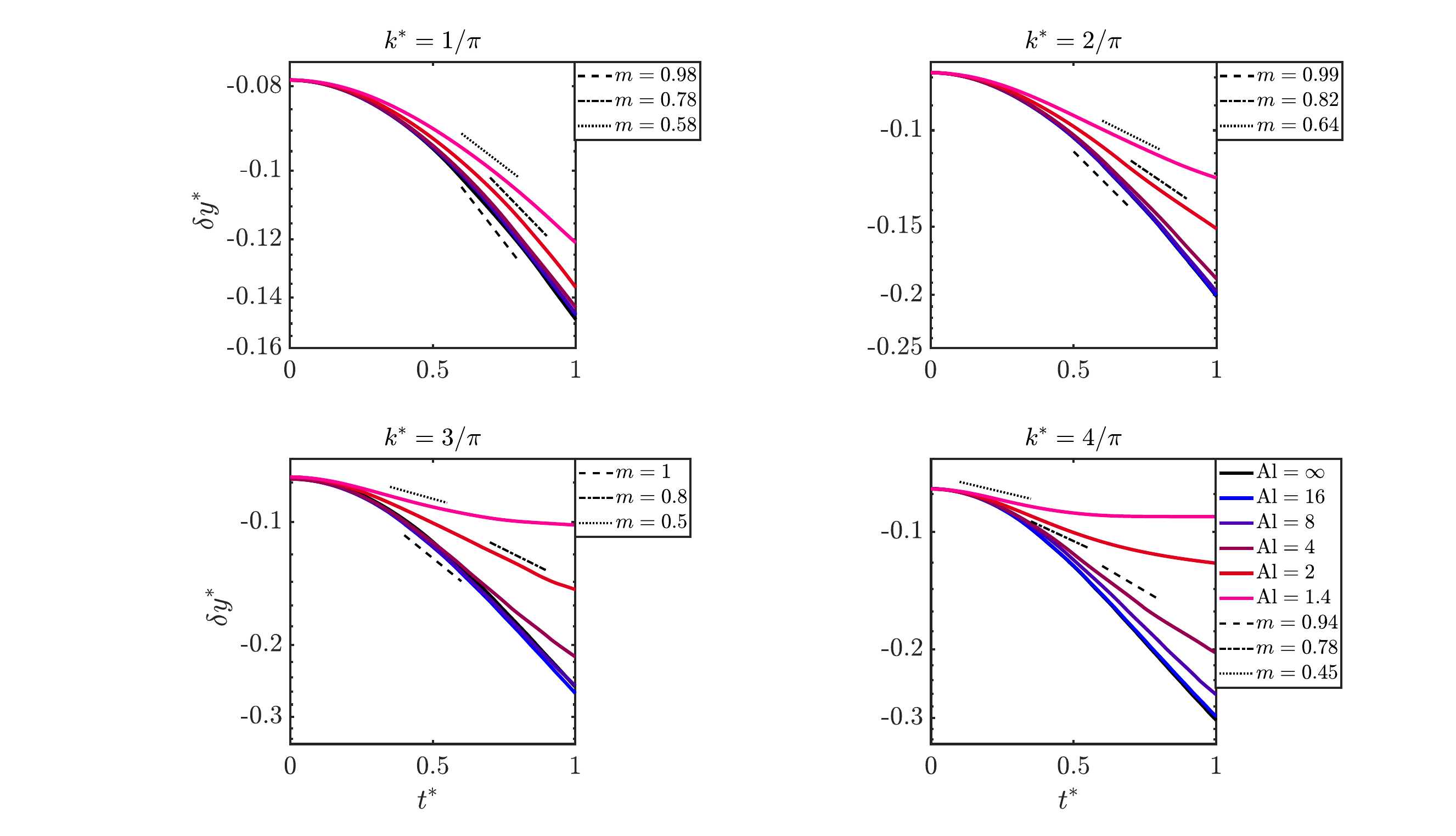}}
% \vspace*{-1cm}
\caption{(\textit{a}) The displacement of the perturbation located at the midpoint of the wavelength is measured during the simulation. (\textit{b}) MRT growth of four different wavenumbers, $k^*=1/\pi$, $2/\pi$, $3/\pi$, and $4/\pi$, for six different Alfv\'en numbers, i.e., $\mathrm{Al}=\infty$, $16$, $8$, $4$, $2$, and $1.4$.}
\label{fig:idealMHD2}
\end{figure}

Using the implemented two-phase ideal MHD solver, MRT instability growth was investigated for five different Alfv\'en numbers of 16, 8, 4, 2, and 1.4, along with the pure hydrodynamic RT case ($\mathrm{Al}=\infty$). The numerical simulations were performed for four different wavenumbers of $k^*=1/\pi$, $2/\pi$, $3/\pi$, and $4/\pi$. The vertical displacement of the perturbation located at the midpoint of the perturbation wavelength (i.e., spike tip), was tracked during the simulation to calculate the growth rate, as shown in Fig.~\ref{fig:idealMHD2}(a). In Fig.~\ref{fig:idealMHD2}(b), this perturbation displacement is plotted as a function of time using a logarithmic scale for the $y-$axis. Therefore, the slope corresponds to the numerical MRT growth rate. 

The dimensionless growth rate is close to unity for all four wavenumbers in the classical RT case, as expected. For each wavenumber, it is observed that decreasing the Alfv\'en number results in a lower growth rate, demonstrating the stabilizing effect of the magnetic field initially present in the lower layer. For $k^*=1/\pi$, the instability growth rates for cases $\mathrm{Al}=16$ and $8$ are quite close (approximately $0.9$), which agrees with the analytical solution (see the $k^*=1/\pi$ line in Fig.~\ref{fig:idealMHD1}b). However, according to the analytical solution, the instability growth rates reduce to $0.78$ and $0.58$ for cases $\mathrm{Al}=2$ and $1.4$, respectively, consistent with the presented numerical results.

For the case $k^*=2/\pi$, the numerical growth rate for $\mathrm{Al}=2$ and $1.4$ are given as $0.82$ and $0.64$, respectively. These values are greater than the corresponding ones for $k^*=1/\pi$, as also inferred from the analytical solution. Moving to $k^*=3/\pi$, the growth rate for the case $\mathrm{Al}=2$ reaches $0.8$, slightly decreasing from the corresponding one for $k^*=2/\pi$, as indicated by Fig.~\ref{fig:idealMHD2}(b). This value further decreases to $0.78$ for the $k^*=4/\pi$ case. Consequently, the numerical solver also captured a decrease in the growth rate for the case $\mathrm{Al}=2$ beyond the wavenumber of $k^*=2/\pi$, similar to the analytical solution. 

For $\mathrm{Al}=1.4$, the numerical growth rates for $k^*=2/\pi$, $3/\pi$, and $4/\pi$ are $0.64$, $0.5$, and $0.45$, respectively, confirming the maximum growth rate for the wavenumber $k^*=2/\pi$. However, the calculated numerical growth rates for the Alfv\'en number of $1.4$ and wavenumbers $3/\pi$ and $4/\pi$ slightly differ from the analytical ones, which are $0.6$ and $0.52$ for $k^*=3/\pi$ and $4/\pi$, respectively. 

Generally, from Fig.~\ref{fig:idealMHD2}(b), it is apparent that for test cases with greater magnetic field density (smaller Alfv\'en number), the numerical solution for the perturbation growth deviates from the analytical one for higher wavenumbers of $k^*=3/\pi$ and $4/\pi$. For example, for the case with an Alfv\'en number and wavenumber of $1.4$ and $4/\pi$, respectively, the perturbation growth differs from the predicted exponential growth of the analytical solution, as can be seen in Fig.~\ref{fig:idealMHD2}(b), especially at the later stages of instability growth. This is explained by the fact that the presented analytical solution is mainly valid in linear regimes for relatively long wavelengths (i.e., small wavenumbers). As the wavenumber value increases and consequently the magnetic tension increases, the accuracy of the analytical solution decreases.

\begin{figure}[]
% \centering
\hspace*{-1.4cm}
{\includegraphics[width = 1.2\textwidth]{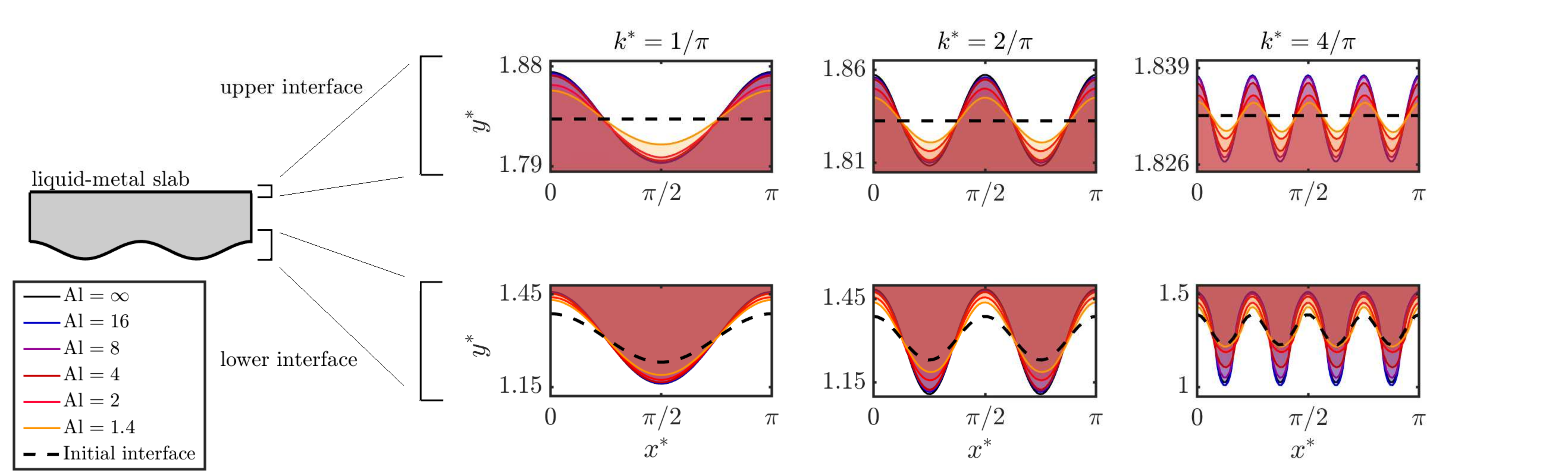}}\\
% \vspace*{-1cm}
\caption{The location of upper and lower interfaces at $t^*=1$ for three wavenumbers, $k^*=1/\pi$, $2/\pi$, and $4/\pi$, across six different Alfv\'en numbers, $\mathrm{Al}=\infty$, $16$, $8$, $4$, $2$, and $1.4$.}
\label{fig:idealMHD3}
\end{figure}

This section concludes with a discussion of the feedthrough effect on the upper interface of the liquid liner. Figure~\ref{fig:idealMHD3} illustrates the location of the upper and lower interfaces at $t^*=1$ across wavenumbers $k^*=1/\pi$, $2/\pi$, and $4/\pi$, for different Alfv\'en numbers. It is evident that for each wavenumber, decreasing the Alfv\'en number leads to a reduction in perturbation growth at the initially unperturbed upper interface. This is due to the fact that the perturbation growth on the upper interface depends on the RT growth of the lower interface, and for each wavenumber decreasing the Alfv\'en number results in reduced RT growth. Additionally, in general, as seen in Fig.~\ref{fig:idealMHD3}, increasing $k^*$ results in a reduction in the feedthrough effect on the upper interface. Consequently, feedthrough is more notable for smaller $k^*$ values, i.e., thinner slabs or smaller wavenumbers. 

\subsection{\label{sec:resistiveMHD} Resistive MHD case}
In the preceding section, the liquid liner was assumed to be perfectly conductive. However, in real-world applications, liquid metals have finite resistivity, which impacts the growth of the MRT instability compared to the ideal case. The effect of magnetic diffusion is examined in the present section. 

\begin{figure}[]
\centering
\subfloat[]{\includegraphics[width = 0.55\textwidth]{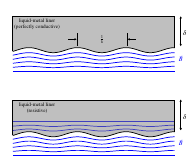}}\\
% \hspace*{-1.8cm}
\subfloat[]{\includegraphics[width = 1.0\textwidth]{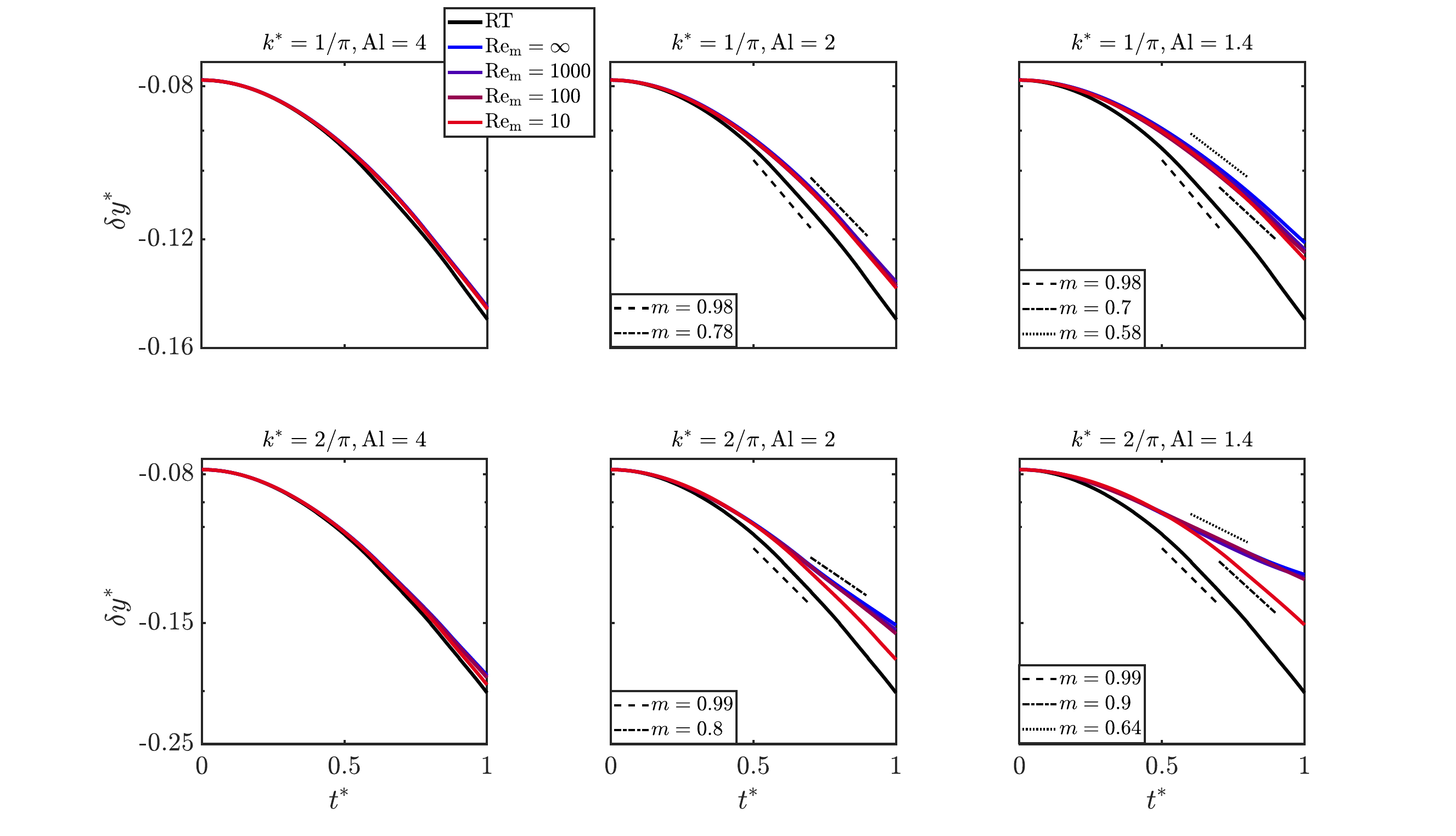}}
% \vspace*{-cm}
\caption{(\textit{a}) The schematic of the problem qualitatively depicting the behavior of magnetic field lines for both ideal and resistive cases. (\textit{b}) MRT growth of two wavenumbers, [top] $k^*=1/\pi$ and [bottom] $k^*=2/\pi$, with three different Alfv\'en numbers of 4, 2, and 1.4, from left to right, for four magnetic Reynolds numbers of $\infty$, $1000$, $100$, and $10$, along with the pure hydrodynamic case.}
\label{fig:resistiveMHD1}
\end{figure}

Figure~\ref{fig:resistiveMHD1}(a) displays the schematic of the problem and qualitatively represents the difference between the configuration of the magnetic field lines for both ideal and resistive cases. Figure~\ref{fig:resistiveMHD1}(b) illustrates the MRT growth rate for three different magnetic Reynolds numbers of 1000, 100, and 10, along with the ideal MHD case, $\mathrm{Re}_\mathrm{m}=\infty$, and their pure hydrodynamic counterpart. The results are depicted for two wavenumbers, $k^*=1/\pi$ and $2/\pi$, and for three different Alfv\'en numbers of $\mathrm{Al}=4$, $2$, and $1.4$, with electrical conductivity ratio set to $\sigma_\mathrm{e,r}=0.1$. Figure~\ref{fig:resistiveMHD1}(b) reveals a negligible difference between the ideal and resistive growth rates for the small wavenumber $k^*=1/\pi$ (top row of Fig.~\ref{fig:resistiveMHD1}b), especially for cases with lower magnetic field values. For instance, a slight difference is observed between the studied cases for $\mathrm{Al}=2$ only towards the end of the simulation time. However, the difference between the ideal and resistive cases becomes more evident for the smallest Alfv\'en number ($\mathrm{Al}=1.4$), and a greater growth rate of $0.7$ for $\mathrm{Re}_\mathrm{m}=10$ is noted compared to the ideal case with the growth rate of $0.58$.

For the cases with $k^*=2/\pi$ (bottom row of Fig.~\ref{fig:resistiveMHD1}b), while the growth rates of the ideal and resistive cases for $\mathrm{Al}=4$ only differ slightly in the final stages of the simulation, the results for smaller Alfv\'en numbers of $2$ and $1.4$ reveal more notable differences between the ideal and resistive cases. In the case of $\mathrm{Al} = 2$, the perturbation growth of the resistive scenario with $\mathrm{Re}_\mathrm{m}=10$ begins to deviate from the ideal case around $t^*=0.6$ and rises to $0.9$. For higher magnetic Reynolds numbers of $1000$ and $100$, the perturbation growth is similar to the ideal case, with only small differences observed around the final time of the simulation. Similar behaviour is detected for $\mathrm{Al}=1.4$, where the growth rate of the resistive case with $\mathrm{Re}_\mathrm{m} = 10$ increases to $0.9$ compared to $0.64$ for the ideal case.  

\begin{figure}[]
% \centering
\hspace*{-1.8cm}
{\includegraphics[width = 1.2\textwidth]{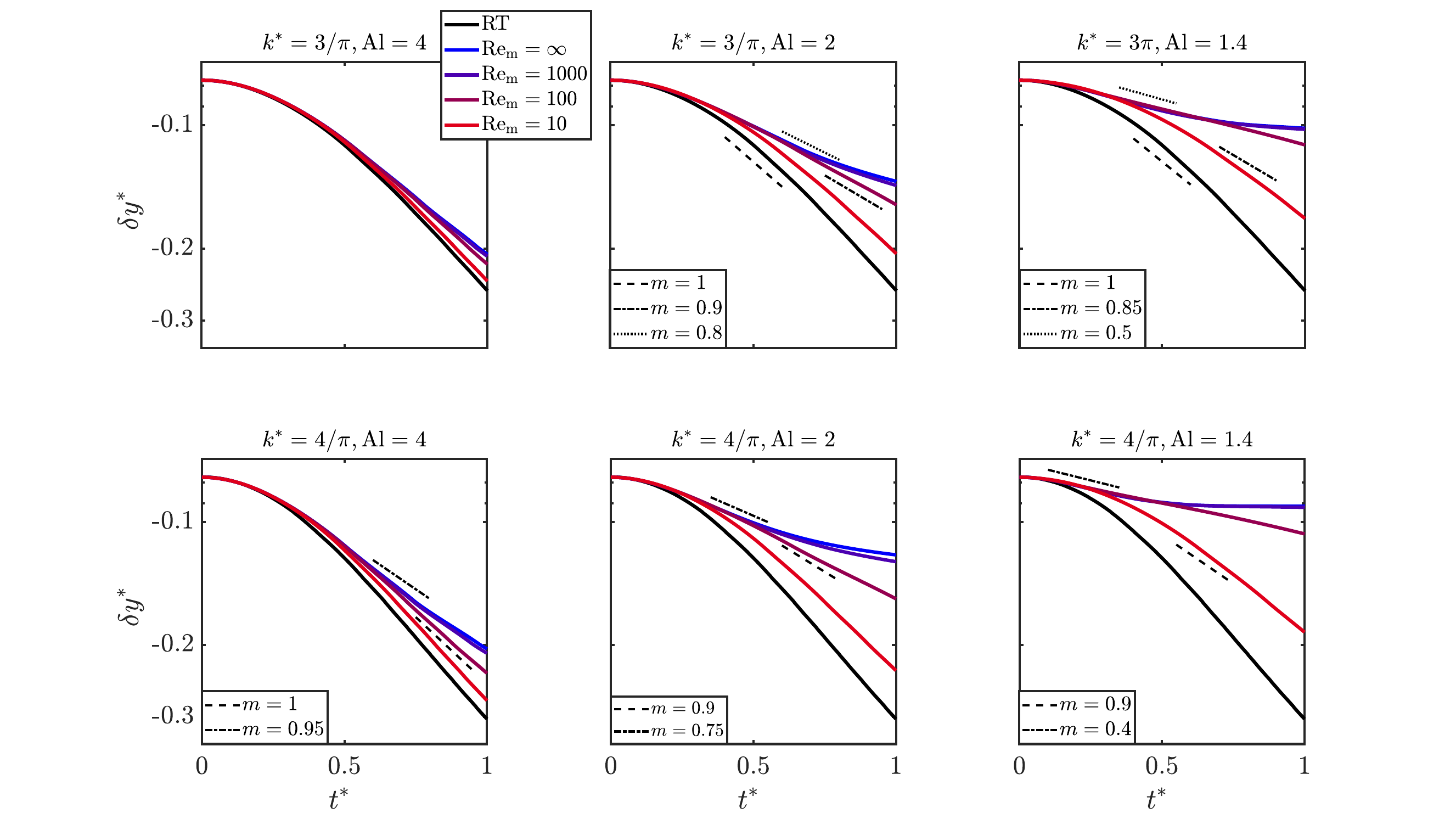}}\\
% \vspace*{-1cm}
\caption{MRT growth of two wavenumbers, [top] $k^*=3/\pi$ and [bottom] $k^* = 4/\pi$, with three different Alfv\'en numbers of 4, 2, and 1.4, from left to right, for four magnetic Reynolds numbers of $\infty$, $1000$, $100$, and $10$, along with the pure hydrodynamic case.}
\label{fig:resistiveMHD2}
\end{figure}

The same results were reproduced for cases with wavenumbers of $k^*=3/\pi$ and $k^*=4/\pi$, shown in Fig.~\ref{fig:resistiveMHD2}. This figure demonstrates that for these higher wavenumbers, even for the high Alfv\'en number ($\mathrm{Al}=4$), the difference between the MRT growth of ideal and resistive cases is more pronounced. Additionally, as displayed in Fig.~\ref{fig:resistiveMHD2}, the magnetic Reynolds number plays an important role in determining the MRT growth rate. For all the presented cases, the MRT growth rate for $\mathrm{Re}_\mathrm{m} = 1000$ is close to the ideal case, as this high magnetic Reynolds number corresponds to a highly conductive medium. However, by decreasing the magnetic Reynolds number, the MRT growth rate increases and becomes closer to that of the pure hydrodynamic case. For example, for an Alfv\'en number of $2$, the growth rate of $\mathrm{Re}_\mathrm{m}=10$ becomes almost similar to that of the pure hydrodynamic case for both wavenumbers (see Fig.~\ref{fig:resistiveMHD2}).  

Based on the conducted numerical simulations, it is concluded that finite resistivity has a more pronounced effect on the MRT growth of an initially magnetic-field-free liquid liner for perturbations with higher wavenumbers, i.e., shorter wavelengths, compared to perturbations with smaller wavenumbers. At higher wavenumbers, the difference between the ideal and resistive cases emerges at an earlier time compared to smaller wavenumbers. Secondly, for smaller Alfv\'en numbers, the distinction between the ideal and resistive cases develops faster, and the impact of magnetic Reynolds number becomes more noticeable. Lastly, the presence of finite resistivity increases the MRT growth rate compared to the ideal case, and for small magnetic Reynolds numbers (i.e., high magnetic diffusivity), the MRT growth rate approaches that of the pure hydrodynamic case. 

\begin{figure}[]
% \centering
\hspace*{-1.5cm}
{\includegraphics[width = 1.2\textwidth]{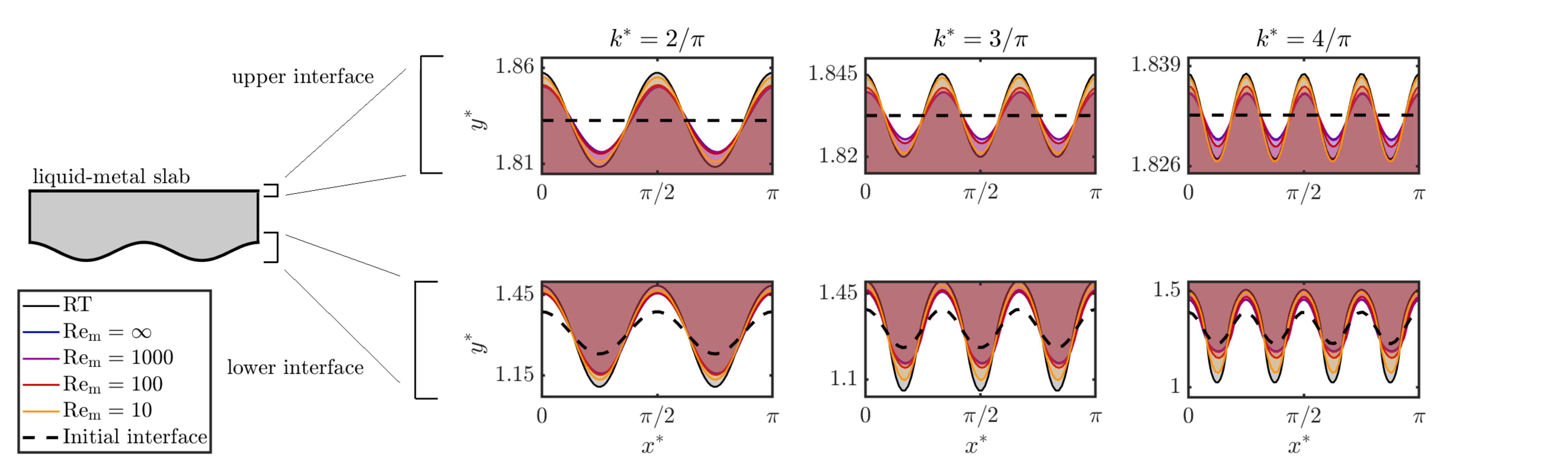}}\\
\caption{The location of upper and lower interfaces at $t^*=1$ for three wavenumbers, $k^* = 2/\pi$, $3/\pi$, and $4/\pi$, with Alfv\'en number of $2$, for four different magnetic Reynolds number of $\infty$, $1000$, $100$, and $10$, along with the hydrodynamic case.}
\label{fig:resistiveMHD3}
\end{figure}

To investigate the potential impact of magnetic diffusion on the morphology of the MRT instability spikes and bubbles, Fig.~\ref{fig:resistiveMHD3} displays the upper and lower interfaces of the liquid liner at $t^*=1$ for three different wavenumbers, $k^*=2/\pi, 3/\pi$, and $4/\pi$, with $\mathrm{Al}=2$. It can be observed that increasing magnetic diffusivity, which corresponds to decreasing the magnetic Reynolds number, leads to the spikes and bubbles of the MRT instability exhibiting growth patterns closer to those observed in pure hydrodynamic cases. Based on the presented results for the upper interface location, it is evident that, as expected, higher $k^*$ values demonstrate less of a feedthrough effect on the upper interface. However, magnetic diffusion increases the feedthrough effect compared to the ideal case, as it enhances the instability growth of the MRT unstable interface.

\begin{figure}[]
\centering
% \hspace*{-2.5cm}
\subfloat[]{\includegraphics[width = 1.0\textwidth]{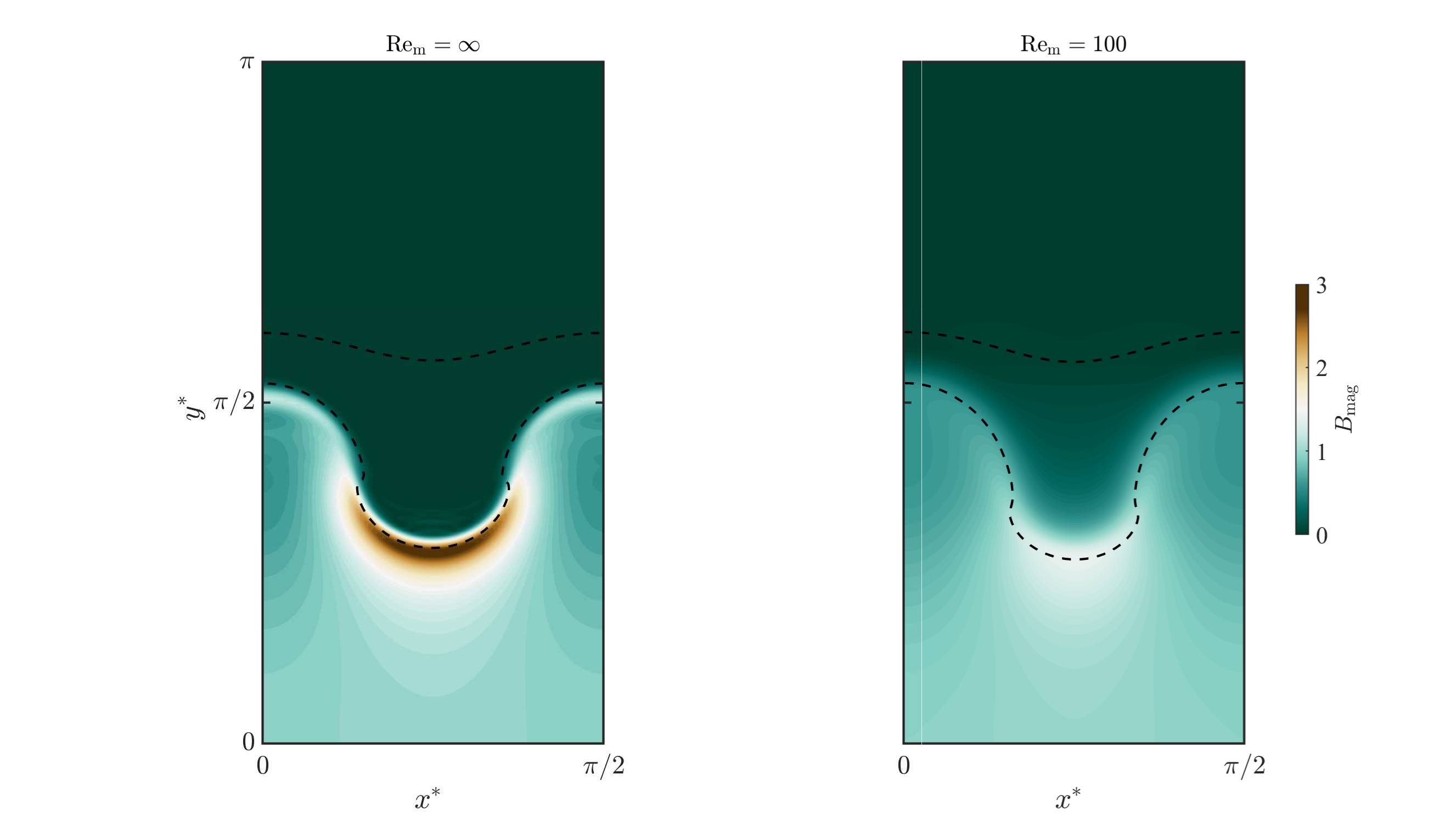}}\\
% \hspace*{-2.5cm}
\subfloat[]{\includegraphics[width = 1.0\textwidth]{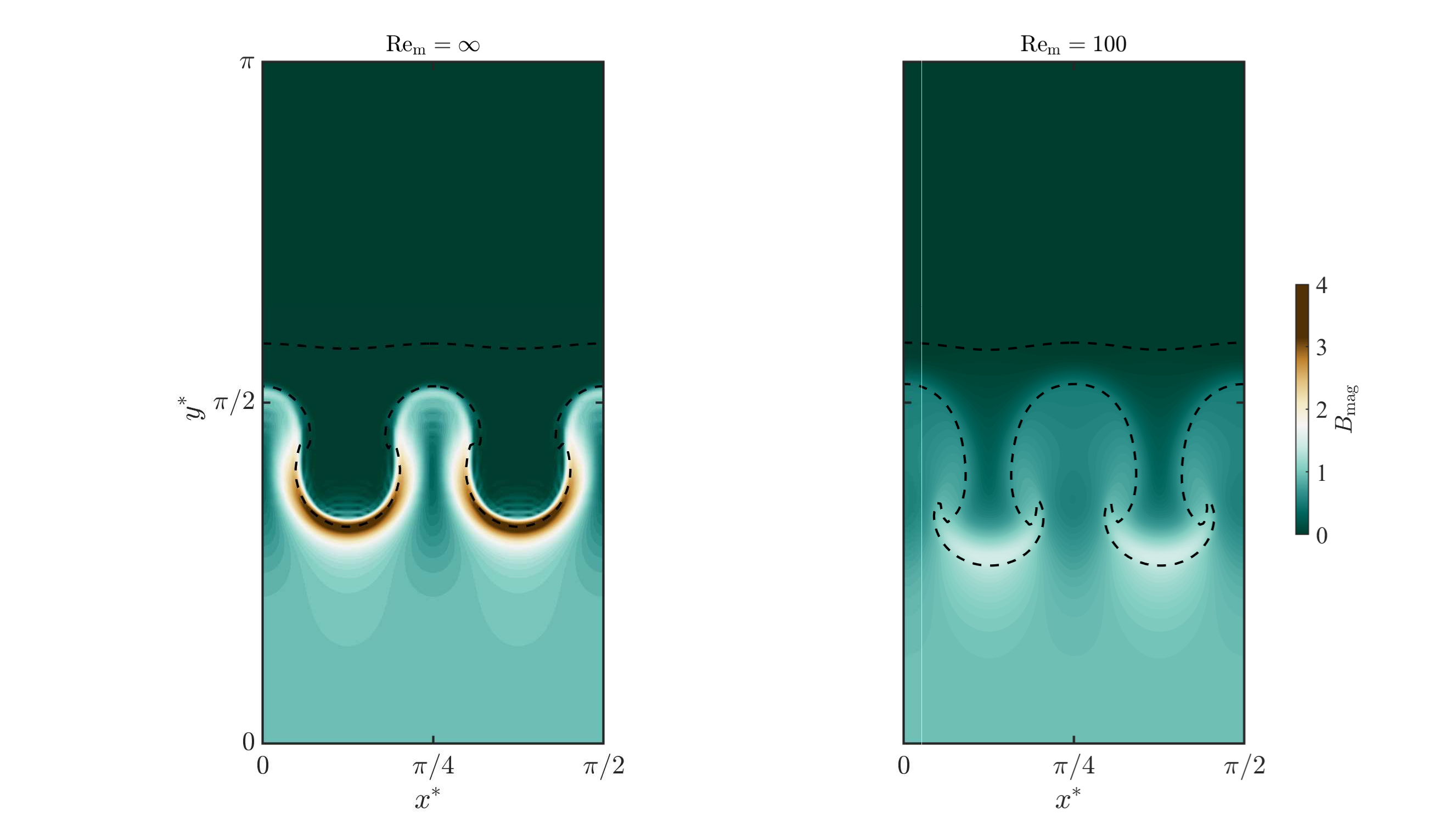}}
\caption{Magnetic field strength at $t^*=2$ for two cases: ideal MHD and resistive MHD, for two wavenumbers: (\textit{a}) $k^*=2/\pi$ and (\textit{b}) $k^*=4/\pi$ with Alfv\'en number of 4. Dashed lines represent the liquid liner interface.}
\label{fig:resistiveMHD4}
\end{figure}

Figure~\ref{fig:resistiveMHD4} depicts the magnetic field contours at $t^*=2$ for both ideal and resistive MHD cases, considering two wavenumbers: $k^*=2/\pi$ and $4/\pi$. The corresponding Alfv\'en and magnetic Reynolds numbers are 4 and 100, respectively. As can be observed from this figure, in the ideal case, no magnetic field has diffused into the liquid liner, whereas in the resistive MHD case, the magnetic field lines have penetrated the liner due to its finite resistivity. In Fig.~\ref{fig:resistiveMHD4}(a), it is apparent that the peak magnetic amplitude in the ideal case is approximately $3$, whereas it is around $1.4$ for the resistive MHD case at $k^*=2/\pi$. This difference confirms that magnetic diffusion results in decreased magnetic compression and, therefore, a reduction in magnetic tension in the resistive MHD case.

The same trend is observed in Fig.~\ref{fig:resistiveMHD4}(b) for the wavenumber $k^*=4/\pi$, with the maximum magnetic field values being almost 4 and 2 for the ideal and resistive MHD cases, respectively. Although the peak magnetic field value in the ideal MHD case is higher for $k^*=4/\pi$ compared to $k^*=2/\pi$, transitioning from the ideal assumption to the resistive MHD case results in reduced magnetic tension and, hence, increased RT growth. 

\begin{figure}[]
% \centering
\hspace*{-2cm}
{\includegraphics[width = 1.2\textwidth]{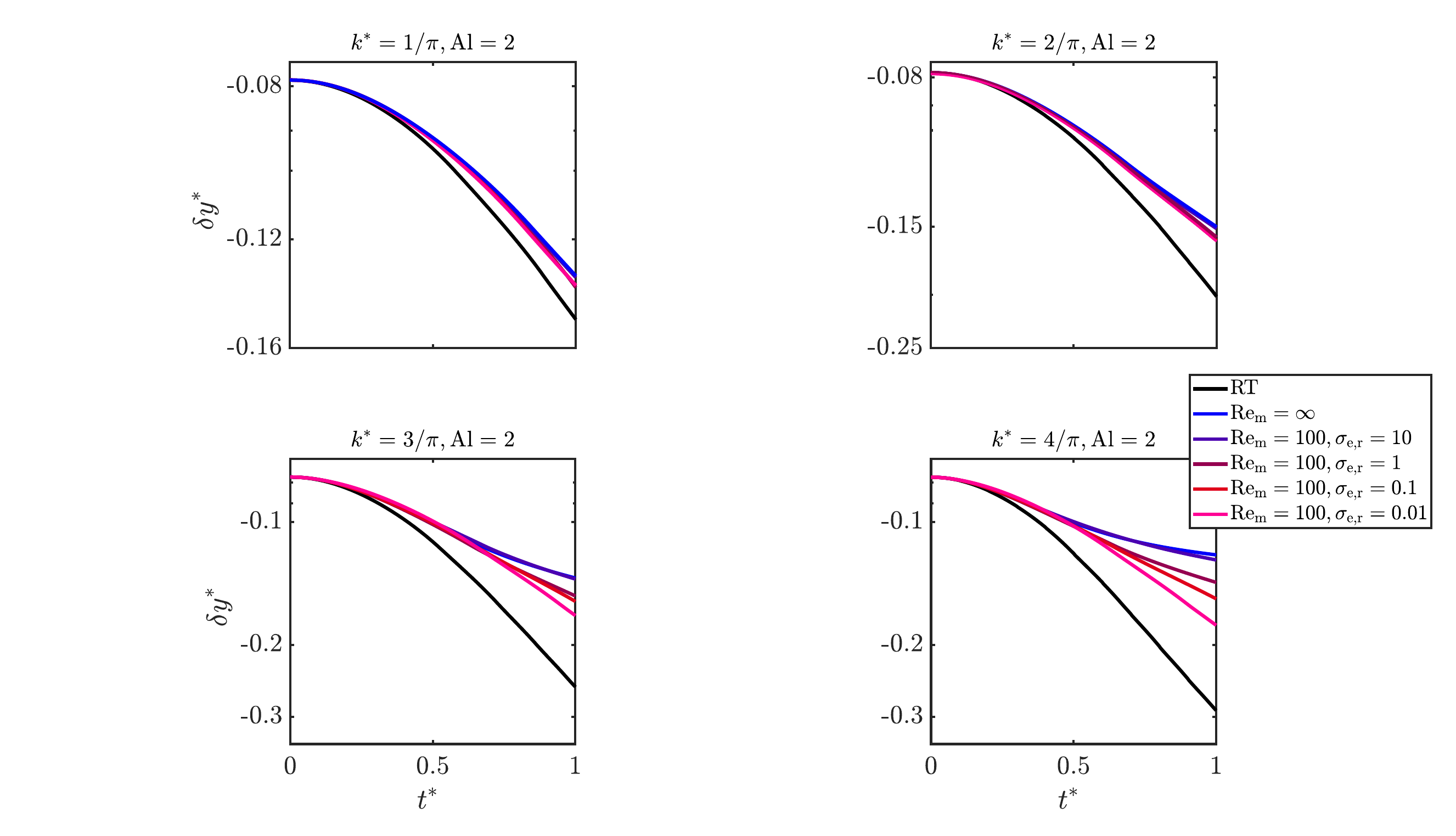}}\\
\caption{MRT growth of four wavenumbers, $k^*=1/\pi$, $2/\pi$, $3/\pi$, and $4/\pi$, with Alfv\'en number and magnetic Reynolds number of 2 and 100, respectively, for four different electrical conductivity ratios of $10$, $1$, $0.1$, and $0.01$.}
\label{fig:resistiveMHD5}
\end{figure}

To conclude this section, the impact of the electrical conductivity ratio across the interface on the perturbation growth rate is investigated. To this end, with Alfv\'en number and magnetic Reynolds number set to 2 and 100, respectively, MRT growth was simulated for four different electrical conductivity ratios of $\sigma_\mathrm{e,r}=10$, $1$, $0.1$, and $0.01$. The results are represented in Fig.~\ref{fig:resistiveMHD5} for wavenumbers $k^*=1/\pi$, $2/\pi$, $3/\pi$, and $4/\pi$. As indicated by this figure, for smaller wavenumbers of $k^*=1/\pi$ and $2/\pi$, the effect of the electrical conductivity ratio on the growth rate is almost negligible, with only slight differences beginning to appear towards the end of the simulation. However, this effect is noticeable for higher wavenumbers, i.e., $k^*=3/\pi$ and $4/\pi$. According to Fig.~\ref{fig:resistiveMHD5}, for the conductivity ratio of 10, meaning the lower region has greater conductivity compared to the liquid liner, the perturbation growth closely resembles that of the ideal MHD case. Nevertheless, for conductivity ratios of $\sigma_\mathrm{e,r} \leq 1$, the MRT growth begins to deviate further from the ideal MHD case. As the electrical conductivity jump across the interface increases, the growth rate tends to approach the classical RT case. This observation could be of significant importance in practical fusion applications, as the electrical conductivity ratio across the liner can become noticeable depending on the conditions. Consequently, the MRT growth rate increases, further diminishing the stabilizing effect of the magnetic field observed in the ideal case. Additionally, as one may expect from the presented results, it was observed that a higher electrical conductivity ratio resulted in an increased feedthrough effect on the upper interface.

% In the following section, the effect of liner surface tension on the instability growth is examined for both ideal and resistive MHD cases.

\subsection{\label{sec:ST} Effect of surface tension}
Surface tension tends to stabilize an interface against the development of RT instability by producing a restoring force, with this force increasing for perturbations with larger wavenumbers. In this section, the numerical solver was utilized to investigate MRT instability growth for both ideal and resistive MHD cases, taking into account the effect of surface tension. Our primary focus is the extent to which the value of the Bond number noticeably influences perturbation growth due to the presence of surface tension. 

According to simulation results, the effect of surface tension on the MRT growth was almost negligible for Bond numbers ranging from $10^7$ to $10^4$, especially for perturbations with smaller wavenumbers. Due to its insignificant impact on the instability growth, the corresponding results are not shown. Therefore, numerical simulations demonstrate that within the Bond number range calculated in Table~\ref{tab:1}, the effect of surface tension appears to be minimal.
\begin{figure}[]
% \centering
\hspace*{-1.0cm}
{\includegraphics[ width = 1.0\textwidth]{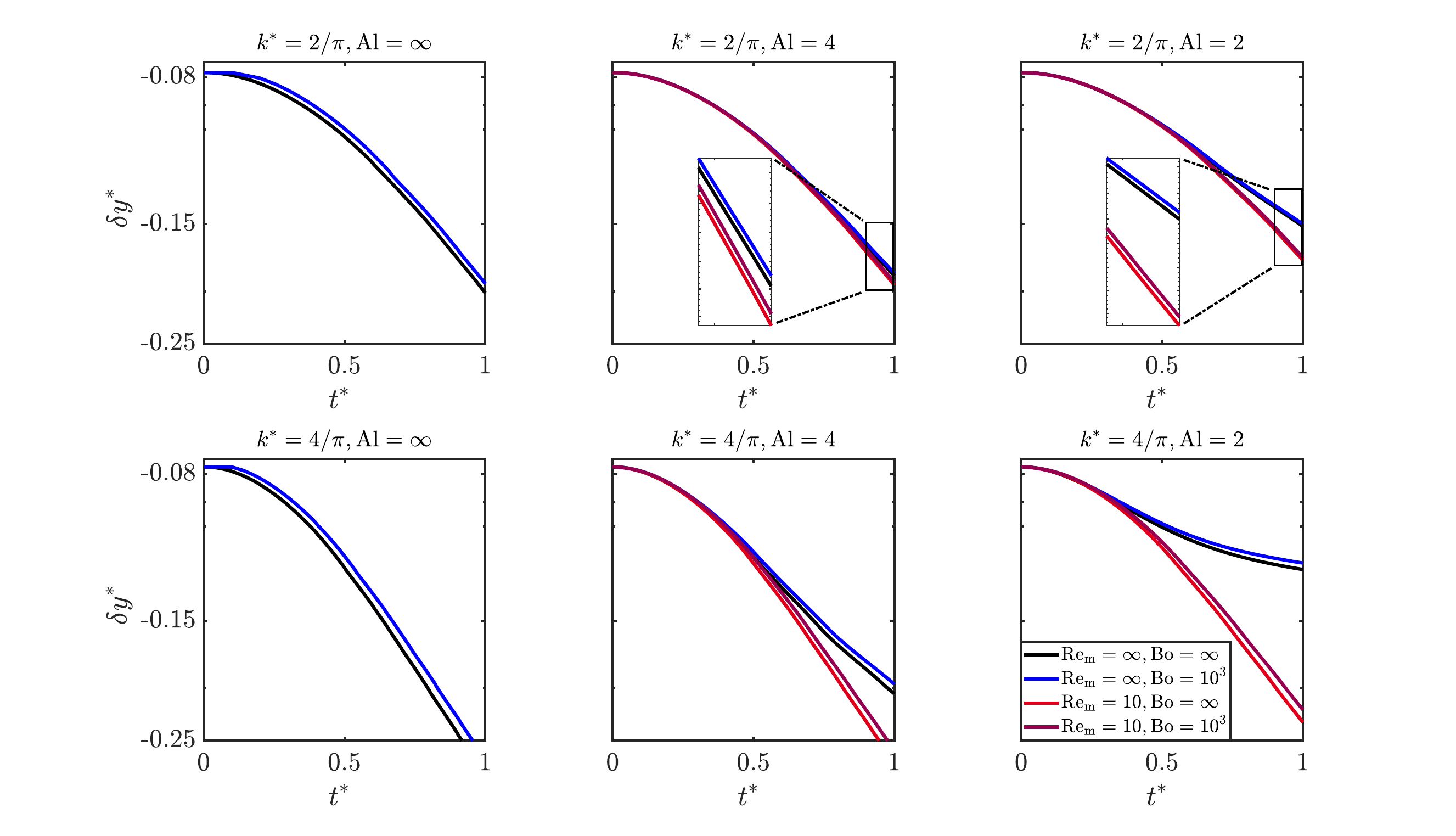}}
\caption{MRT growth of two wavenumbers [top] $k^*=2/\pi$ and [bottom] $k^*=4/\pi$ for three Alfv\'en numbers, $\infty$, 4, and 2, with a Bond number set to $\mathrm{Bo}=10^3$, for both ideal and resistive MHD cases.}
\label{fig:ST1}
\end{figure}

The simulation results for the Bond number of $\mathrm{Bo}=10^3$ are presented in Fig.~\ref{fig:ST1} for two wavenumbers of $k^*=2/\pi$ and $4/\pi$. This figure illustrates the results for Alfv\'en numbers of $\mathrm{Al}=\infty$ (classical RT), $4$, and $2$ with two magnetic Reynolds numbers of $\mathrm{Re}_\mathrm{m}=\infty$ (ideal MHD case) and $10$ and with an electrical conductivity ratio of $0.1$. As depicted in Fig.~\ref{fig:ST1}, the stabilization effect of surface tension is more pronounced for the higher wavenumber of $k^*=4/\pi$ compared to $k^*=2/\pi$. However, even for the case of $k^*=2/\pi$, it can be observed that considering the effect of surface tension has led to a smaller growth rate for both ideal and resistive cases (see the insets of Fig.~\ref{fig:ST1}(top row)).

 As seen in Fig.~\ref{fig:ST1}(bottom row), with the stabilization effect of surface tension being more pronounced in the case of $k^* = 4/\pi$, it becomes apparent that, for the ideal case, regardless of the Alfv\'en number value, the presence of surface tension leads to reduced perturbation growth. The same behaviour is evident for the resistive case, where the instability growth has decreased for $\mathrm{Bo}=10^3$ compared to the case of $\mathrm{Bo}=\infty$. Based on the presented results, it can be concluded that the stabilizing effect of surface tension is preserved for both ideal and resistive MHD cases. Furthermore, upon examining the evolution of the upper interface, we noted that accounting for surface tension leads to a reduced feedthrough effect on the upper interface. This reduced feedthrough effect is not only due to the reduced RT instability growth at the lower interface, but also the presence of surface tension at the upper interface serves to dampen the instabilities forming at that surface. The visual presentation of surface tension impact on feedthrough at the upper interface is further elaborated in the following section.

\section{\label{sec:discussion} Discussion}
A discussion of the numerical results and analysis presented in Sec.~\ref{sec:results} is provided here, along with an explanation of the underlying physics.

The analytical and numerical results of the MRT instability growth in the ideal MHD scenario suggest that the presence of a horizontal magnetic field in the lower layer reduces the RT instability growth of a liquid-metal liner. This reduction in MRT growth can be explained by the frozen-in law, stating that in ideal MHD flows, magnetic field lines
are attached to the velocity field. Therefore, as the instability ripples begin to grow, magnetic field lines trapped in the lower layer also start to ripple and bend (see Fig.~\ref{fig:idealMHD1}a). Consequently, the bent magnetic field lines experience tension. The resulting restoring force induced in magnetic field lines due to tension acts as a source of energy sink, thereby decreasing the MRT growth rate.
 
Numerical results mainly deviate from the analytical solution for smaller Alfv\'en numbers and higher wavenumbers. This deviation is attributed to the limitation of the analytical solution, which is primarily valid in linear regimes for relatively long wavelengths (i.e., small wavenumbers) and small magnetic tension values. As the wavenumber value increases, and consequently, the magnetic tension becomes stronger, the accuracy of the analytical solution decreases, as noted by Weis~\cite{weis2015thesis}. Another constraint of the analytical solution is its failure to account for the effect of finite resistivity on MRT growth, highlighting the importance of a numerical solver in exploring the interplay between magnetic tension and magnetic diffusion across different instability development regimes.

\begin{figure}[]
\centering
\hspace*{-2.5cm}
\subfloat[]{\includegraphics[width = 1.2\textwidth]{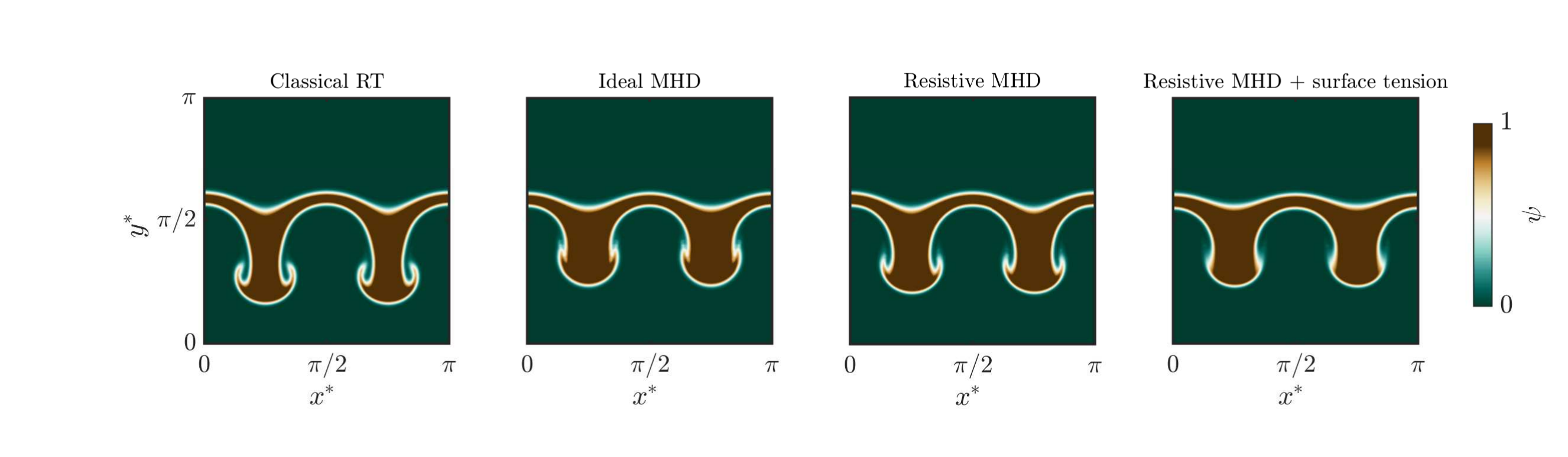}}\\
\hspace*{-2.5cm}
\subfloat[]{\includegraphics[width = 1.2\textwidth]{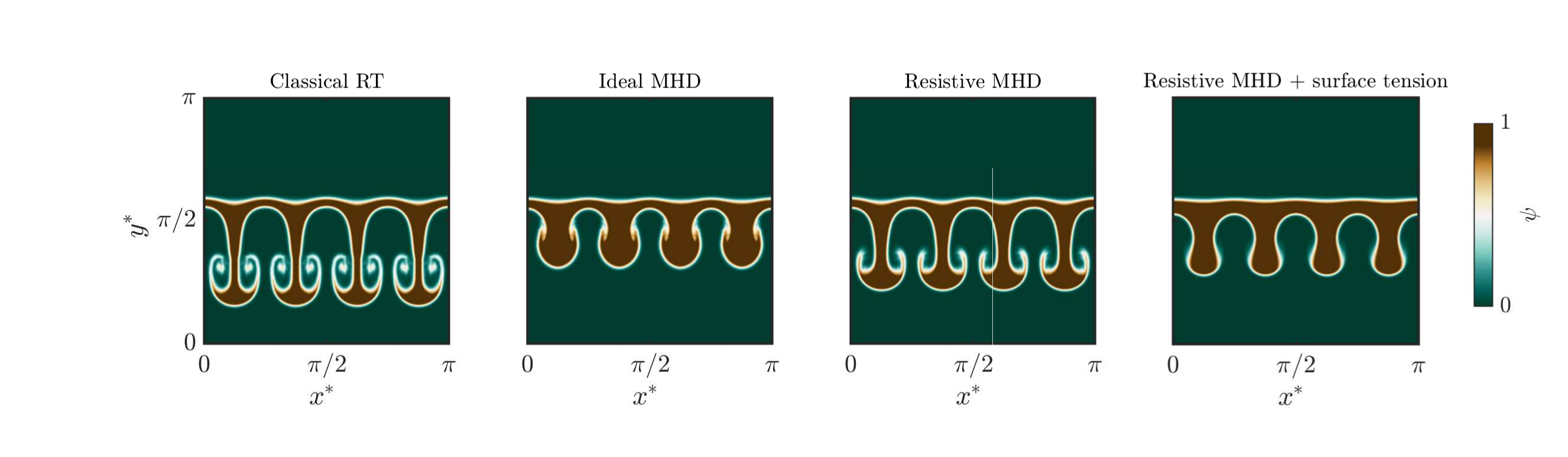}}
\caption{Liquid liner interface at $t^*=2.5$ for four cases: pure hydrodynamic, ideal MHD ($\mathrm{Al}=4$), and resistive MHD ($\mathrm{Al}=4$, $\mathrm{Re}_\mathrm{m}=100$), without and with surface tension, from left to right, for two wavenumbers (\textit{a}) $k^*=2/\pi$ and (\textit{b}) $k^*=4/\pi$.}
\label{fig:conclusion}
\end{figure}

Figure~\ref{fig:conclusion} illustrates the liquid liner interface at $t^*=2.5$ for four different cases: pure hydrodynamic, ideal MHD with $\mathrm{Al}=4$, resistive MHD with a magnetic Reynolds number of $\mathrm{Re}=100$, and resistive MHD case with surface tension (Bond number assumed to be 100). Results are depicted for two wavenumbers: $k^*=2/\pi$ (Fig.~\ref{fig:conclusion}a) and $k^*=4/\pi$ (Fig.~\ref{fig:conclusion}b). The ideal MHD case for the two studied wavenumbers indicates that the stabilizing effect of magnetic tension is more pronounced for higher wavenumbers. One may ascribe this behaviour to the induced tension in the magnetic field during the MRT growth. Through the development of the MRT instability, the bending of magnetic field lines generates Alfv\'en waves, and the magnetic tension part of the Lorentz force is proportional to $\left(\boldsymbol{k} \cdot \boldsymbol{v}_\text{Al} \right)^2$. Hence, as the wavenumber increases, the stabilizing effect of the magnetic field due to the greater magnetic tension becomes more pronounced, leading to further suppression of instabilities at shorter wavelengths. The schematic presented in Fig.~\ref{fig:idealMHD1}(a) demonstrates that for shorter wavelengths, the curvature of magnetic field lines is greater, resulting in larger induced magnetic tension.

The inclusion of magnetic diffusion leads to increased instability growth for both wavenumbers, as displayed in Fig~\ref{fig:conclusion}, compared to the ideal MHD case. In the presence of finite resistivity, the magnetic field lines are not confined within the boundaries of their corresponding medium (i.e., frozen-in law) any longer due to magnetic diffusion. Therefore, magnetic field lines can slip out of the lower region and diffuse into the liquid liner as shown in Fig.~\ref{fig:resistiveMHD1}(a), thereby reducing the restoring force and the stabilization effect compared to the ideal case. Thus, it can be concluded that ideal MHD analysis may significantly overestimate the stabilizing effect of the magnetic field.

As indicated by Fig.~\ref{fig:conclusion}, finite resistivity has a more pronounced effect on the MRT growth of an initially magnetic-field-free liquid liner for perturbations with higher wavenumbers (i.e., shorter wavelengths). This finding is analogous to that of Sun \textit{et al.}~\cite{sun2023rayleigh}, who investigated the effect of magnetic diffusion on the MRT growth for a single interface geometry with a constant magnetic field present in both heavy and light fluids.

In Sec.~\ref{sec:resistiveMHD}, it became evident that increasing magnetic diffusivity, which corresponds to decreasing the magnetic Reynolds number, leads to the spikes and bubbles of the MRT instability exhibiting growth patterns closer to those observed in pure hydrodynamic cases. This observation aligns with the findings of Samulski \textit{et al.}~\cite{samulski2022deceleration}, who investigated MRT instability growth during the deceleration phase of ICF implosion with a constant magnetic field imposed in the domain. According to their study, the observed MRT growth and interface morphology for the resistive MHD case closely resembled the hydrodynamic case, with only slight differences noted~\citep{samulski2022deceleration}. 

The stabilizing mechanism of surface tension, generating restoring forces, is akin to that in the MRT case. Hence, in the literature, the effect of a magnetic field in the ideal MHD case is often considered analogous to the presence of surface tension. However, our numerical simulation showed that for high Bond numbers, the effect of surface tension is almost negligible. In Fig.~\ref{fig:conclusion}, for $\mathrm{Bo}=100$, the liquid-metal interface is noticeably stabilized, especially for the higher wavenumber, $k^*=4/\pi$. Furthermore, it is visually evident that considering the surface tension effect has led to smaller perturbation growth at the upper interface.

In general, the results exhibit that increasing the wavenumber, $k^*$, reduces the feedthrough effect. Previous studies by Lau \textit{et al.}~\cite{lau2011anisotropy}, Weis \textit{et al.}~\cite{weis2014temporal}, and Weis~\cite{weis2015thesis} have also reported that increasing $k^*$ lessens feedthrough, and for $k^* \gg 1$, the feedthrough effect becomes virtually negligible. This observation can also be justified based on the reported feedthrough factor by Taylor~\cite{Taylor1950} for a liquid slab of finite thickness, $\delta$. Taylor~\cite{Taylor1950} showed that the amplitude of ripples on the RT stable surface of the liquid layer grows a factor of $e^{-k \delta}$ less than the perturbations on the RT unstable interface. Consequently, for larger $k^*$ values, interpreted as thicker slabs or higher wavenumbers, the feedthrough effect on the upper interface is smaller. 

\begin{figure}[]
\centering
% \hspace*{-1.5cm}
% \subfloat[]{\includegraphics[width = 0.8\textwidth]{figures/k_2_pi-1.png}}\\
% \vspace*{1.5cm}
{\includegraphics[width = 0.9\textwidth]{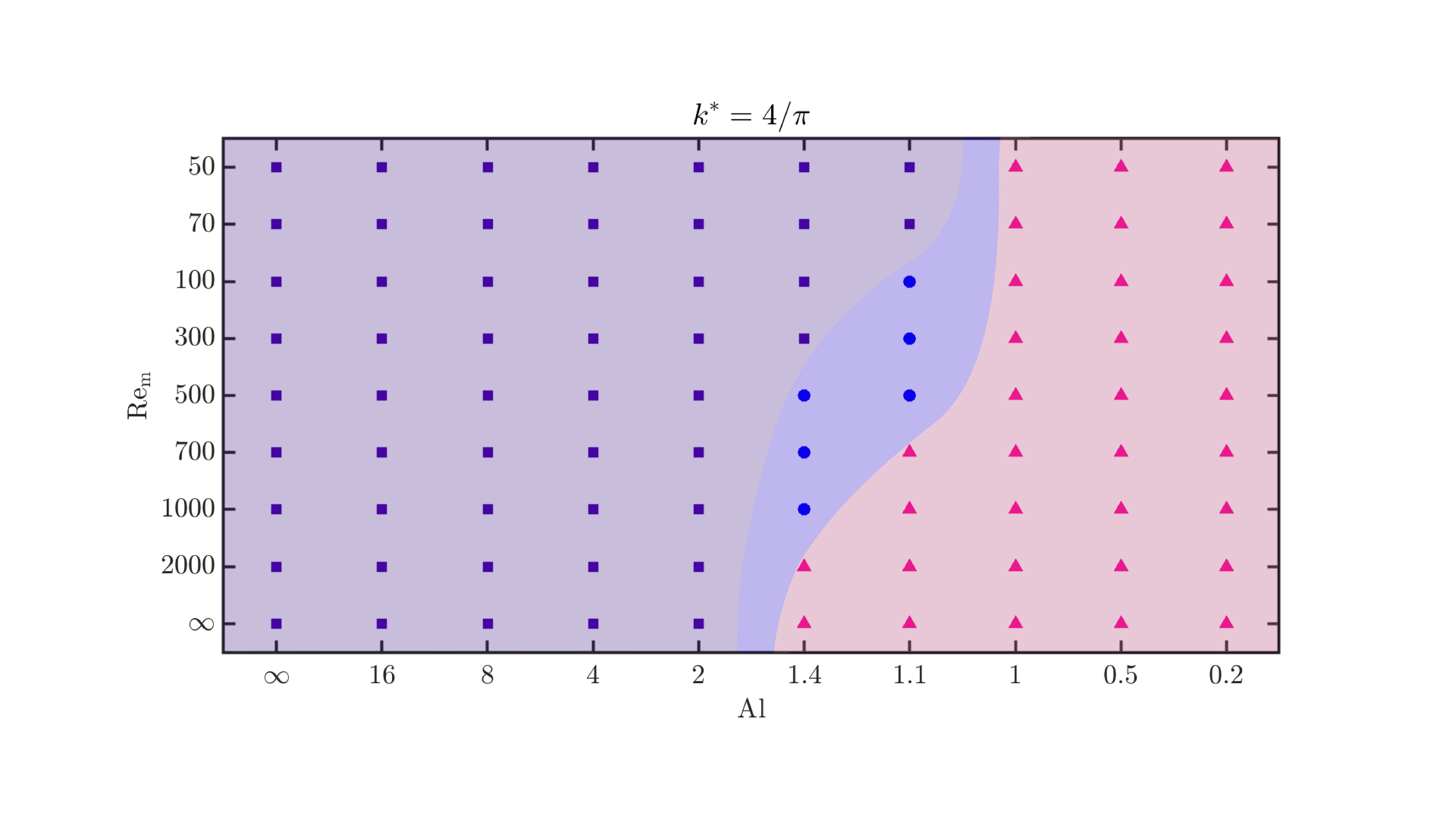}}
\caption{Stability analysis for $k^*=4/\pi$ across various Alfv\'en numbers and magnetic Reynolds numbers, with an electrical conductivity ratio of 0.1. Purple (\emptysquare{mypurple}), pink (\emptytriangle{mypink}), and blue ($\color{blue}{\bullet}$) regions indicate unstable, stable, and transition from initially unstable to stable cases, respectively.}
\label{fig:map}
\end{figure}

The interplay between magnetic tension and magnetic diffusion can be studied across different regimes and classified as a map, which is a function of the two governing parameters, i.e., the Alfv\'en number and the magnetic Reynolds number. Using the established numerical toolkit, this map is represented in Fig.~\ref{fig:map} for the wavenumbers $k^*=4/\pi$.

According to this figure, three distinct behaviours are observed. The purple region (data shown with \emptysquare{mypurple}) illustrates an unstable region in which the RT instability develops. This region mostly corresponds to higher Alfv\'en numbers. In smaller Alfv\'en regimes, the initially perturbed interface starts to oscillate in time, and the interface is observed to be RT stable, indicated in pink (\emptytriangle{mypink}). One important observation is the effect of the magnetic Reynolds number on causing the stable case to become unstable. For instance, for wavenumber $k^*=4/\pi$, in the ideal MHD case ($\mathrm{Re}_\mathrm{m}=\infty$), the RT becomes stable for the Alfv\'en number of 1.4. However, decreasing the Reynolds number to $\mathrm{Re}_\mathrm{m}=50$, the instability starts to grow in time.  The region shown in blue (with markers $\color{blue}{\bullet}$) corresponds to the regime where the perturbation starts to grow; however, after some time, it begins to oscillate, due to an increase in magnetic tension, which has a stabilizing effect on the MRT growth.

\section{\label{sec:conclusion} Conclusion}
This study numerically investigates the MRT instability growth and feedthrough in an initially magnetic-field-free liquid-metal liner, with an axial magnetic field of arbitrary magnitude imposed in the lower layer. To this end, a novel second-order numerical solver was introduced for modelling two-phase incompressible MHD flows within the finite-difference framework.
The MRT instability growth was analyzed for both scenarios of perfectly conducting and resistive liners, and the effect of the Alfv\'en number and magnetic Reynolds number was examined.

The results indicate that magnetic diffusion primarily affects the MRT growth rate for higher wavenumbers, while for smaller wavenumbers, the effect of finite resistivity is observed over a longer duration of instability development. Furthermore, it is demonstrated that decreasing the Alfv\'en number results in the faster emergence of the magnetic diffusion effect on the MRT growth. Additionally, a greater electrical conductivity jump across the liner leads to increased perturbation growth. Lastly, the surface tension effect is shown to be negligible for high Bond numbers, while for relatively smaller values of the Bond number, the stabilizing effect of surface tension is observed for both ideal and resistive MHD cases, particularly notable for higher wavenumbers.

% The results suggest that magnetic diffusion adversely affects the stabilizing effect of the magnetic tension, especially for higher wavenumbers. 

% \begin{acknowledgments}
% The authors would like to thank Ivan Khalzov for his valuable comments and suggestions. 
% \end{acknowledgments}

% \section*{DATA AVAILABILITY}
% The data that support the findings of this study are available
% from the corresponding author upon reasonable request.

\appendix

\section{\label{appendixA} Evaluation of two-phase incompressible MHD numerical solver}
The single interface Rayleigh--Taylor instability with a constant horizontal magnetic field, $\left(B_x, 0, 0 \right)$, in both liquid and gas phases was studied as a benchmark to evaluate the accuracy, convergence, and performance of the implemented two-phase incompressible MHD solver. In this test case, a two-dimensional rectangular domain $[x,y] \in [0,1] \times [0,4]$, with a fluid phase filling the top half of the domain, was considered. The fluid interface at $y_0=2$ was initialized with a small sinusoidal perturbation with the wavelength and amplitude of $2\pi$ and $0.1$, respectively. The density values were set to $\rho_\mathrm{g}=1$ and $\rho_\mathrm{l}=3$ with the gravity acting downwards with a magnitude of unity. The initial magnetic field value and magnetic permeability are set to $B_x=0.1$ and $\mu_\mathrm{m}=1$, respectively. The simulation was performed for four different grid resolutions of $16 \times 64$, $32 \times 128$, $64 \times 256$, and $128 \times 512$ with a constant time step of $\Delta t= 5 \times 10^{-4}/\sqrt{\mathrm{At}}$.

\begin{figure}[]
% \centering
\hspace*{0.4cm}
\subfloat[]{\includegraphics[width = 0.9\textwidth]{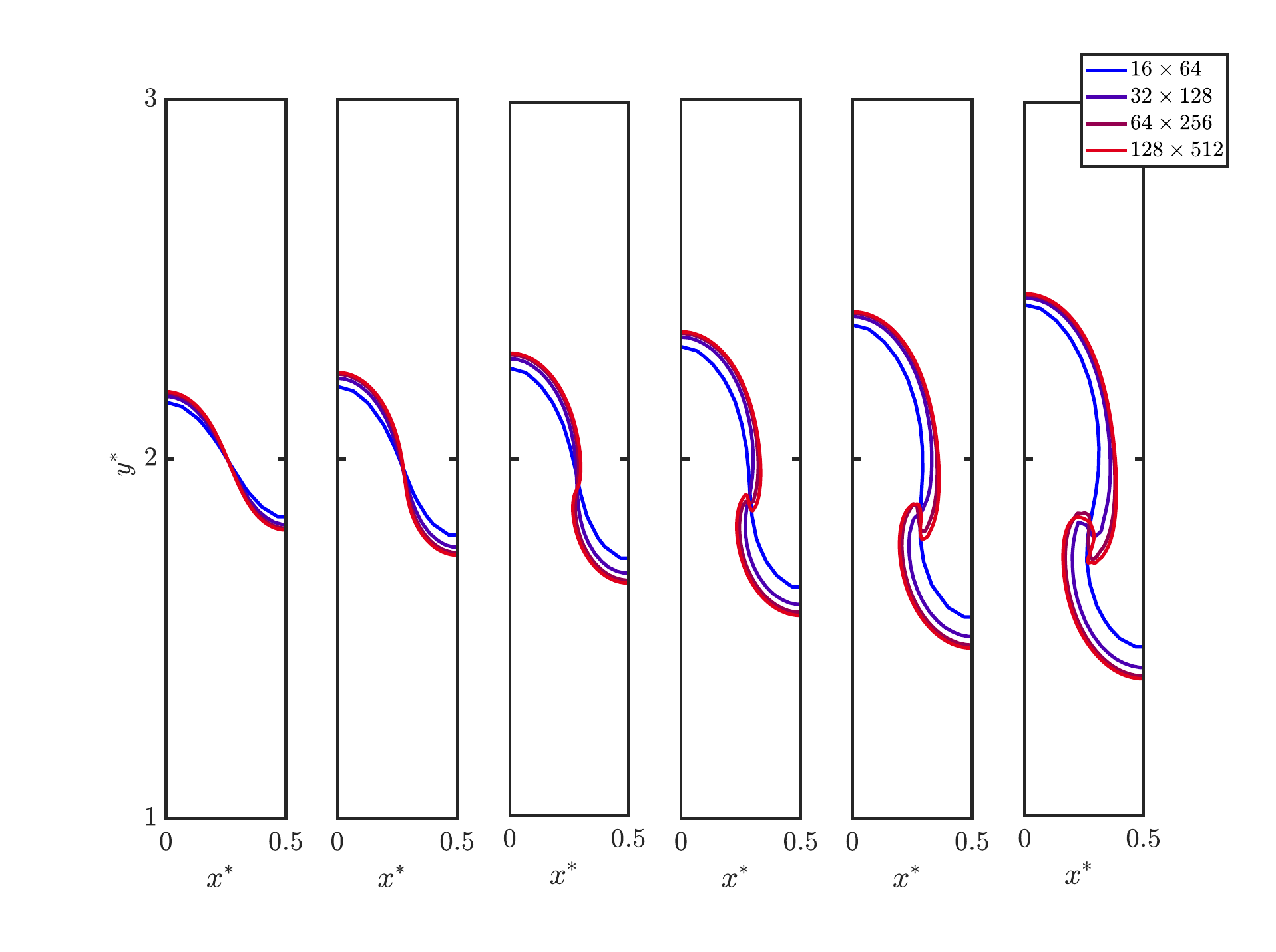}}\\
\hspace*{-2.4cm}
\subfloat[]{\includegraphics[width = 0.7\textwidth]{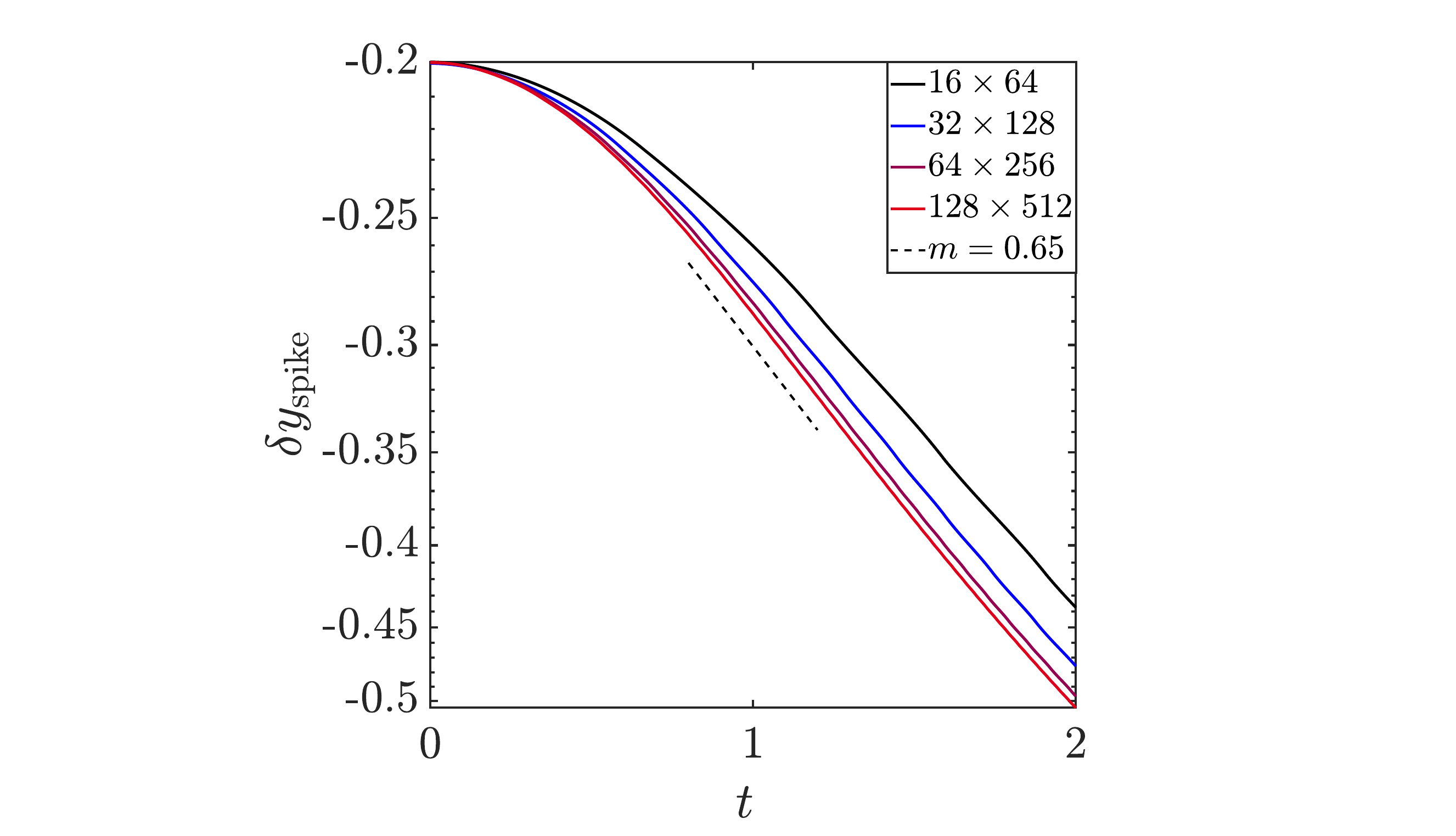}}
\hspace*{-1cm}
\subfloat[]{\includegraphics[width = 0.7\textwidth]{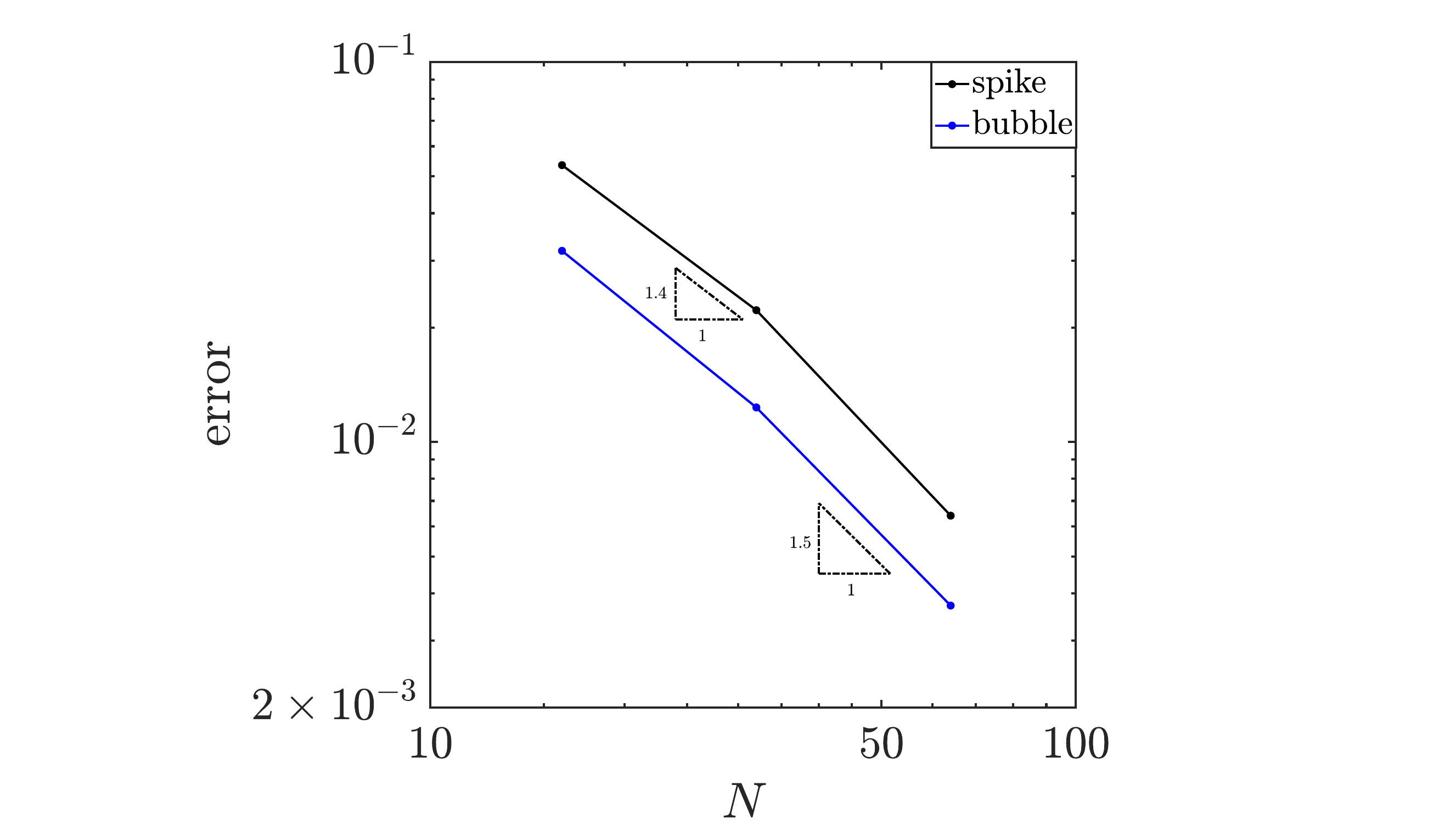}}
\caption{(\textit{a}) The interface of the MRT instability with the density ratio of $3$ at $t \sqrt{\mathrm{At}}=0.75, 1, 1.25, 1.5, 1.75$, and $2$, from left to right, for four different mesh resolutions. (\textit{b}) MRT growth rate for four different mesh resolutions. (\textit{c}) Order of accuracy analysis for the implemented solver.}
\label{fig:Appendix1}
\end{figure}

Figure~\ref{fig:Appendix1}(a) displays the results for the four different mesh resolutions at time $t \sqrt{\mathrm{At}}=0.75$ to $2$ with an increment of 0.25. As depicted in this figure, refining the mesh leads to a more accurate representation of the RT instability features, and for mesh resolutions of $64 \times 256$ and $128 \times 512$, the numerical results are closely matched. For a more thorough quantitative analysis, we compared the obtained numerical growth rate with the analytical solution, shown in Fig.~\ref{fig:Appendix1}(b). The analytical growth rate is calculated as \citep{samulski2022deceleration}
\begin{equation}
    \omega^2 = g k \mathrm{At} - \frac{B^2 k^2}{\mu_\mathrm{m} (\rho_1 + \rho_2)},
\end{equation}
which predicts a growth rate of $0.7$ for this test case. From Fig.~\ref{fig:Appendix1}(b), it is evident that as the grid resolution increases, the numerical growth rate converges to $0.65$, closely matching the analytical solution.

Assuming the solution of the finest mesh, 128 × 512, as an analytical solution, the convergence
rate of the numerical solver is computed. To this end, the $L_2$ norm of the spike and bubble locations during the simulation is calculated. As indicated by Fig.~\ref{fig:Appendix1}(c), the obtained convergence rate is around 1.5, which is close to the expected second order, confirming the solver's accuracy and robustness.

% \nocite{*}
% \bibliography{aipsamp}% Produces the bibliography via BibTeX.

\newpage

\bibliographystyle{elsarticle-num.bst}  %bibliographic style for numerical references
\bibliography{main}

\end{document}